\documentclass[fleqn,usenatbib]{mnras}

\usepackage{newtxtext,newtxmath}

\usepackage[T1]{fontenc}

\DeclareRobustCommand{\VAN}[3]{#2}
\let\VANthebibliography\thebibliography
\def\thebibliography{\DeclareRobustCommand{\VAN}[3]{##3}\VANthebibliography}

\usepackage{graphicx}	
\usepackage{amsmath,bm}	

\usepackage[dvipsnames]{xcolor}

\usepackage[normalem]{ulem}

\newcommand{\corr}[1]{{#1}}

\title[Matching Bayesian and frequentist coverage probabilities]{Matching Bayesian and frequentist coverage probabilities when using an approximate data covariance matrix}

\author[Will J. Percival et al.]{
  Will J. Percival$^{1,2,3}$\thanks{E-mail: will.percival@uwaterloo.ca},
  Oliver Friedrich$^{4,5}$,
  Elena Sellentin$^{6,7}$,
  Alan Heavens$^{8}$
\\
$^{1}$ Waterloo Centre for Astrophysics, University of Waterloo, Waterloo, ON N2L 3G1, Canada \\ 
$^{2}$ Department of Physics and Astronomy, University of Waterloo, Waterloo, ON N2L 3G1, Canada \\
$^{3}$ Perimeter Institute for Theoretical Physics, 31 Caroline St. North, Waterloo, ON N2L 2Y5, Canada \\
$^{4}$ Kavli Institute for Cosmology, University of Cambridge, CB3 0HA Cambridge, United Kingdom \\
$^{5}$ Churchill College, University of Cambridge, CB3 0DS Cambridge, United Kingdom\\
$^{6}$ Mathematical Institute, Leiden University, Snellius Gebouw, Niels Bohrweg 1, NL-2333 CA Leiden, The Netherlands\\
$^{7}$ Leiden Observatory, Leiden University, Oort Gebouw, Niels Bohrweg 2, NL-2333 CA Leiden, The Netherlands\\
$^{8}$Imperial Centre for Inference and Cosmology (ICIC), Department of Physics, Imperial College London, Blackett Laboratory,\\
Prince Consort Road, London SW7 2AZ, UK
}

\date{Accepted XXX. Received YYY; in original form ZZZ}

\pubyear{2021}

\begin{document}
\label{firstpage}
\pagerange{\pageref{firstpage}--\pageref{lastpage}}
\maketitle

\begin{abstract}
  Observational astrophysics consists of making inferences about the Universe by comparing data and models. The credible intervals placed on model parameters are often as important as the maximum {\em a posteriori} probability values, as the intervals indicate concordance or discordance between models and with measurements from other data. Intermediate statistics (e.g. the power spectrum) are usually measured and inferences made by fitting models to these rather than the raw data, assuming that the likelihood for these statistics has multivariate Gaussian form. The covariance matrix used to calculate the likelihood is often estimated from simulations, such that it is itself a random variable. This is a standard problem in Bayesian statistics, which requires a prior to be placed on the true model parameters and covariance matrix, influencing the joint posterior distribution. As an alternative to the commonly-used independence Jeffreys prior, we introduce a prior that leads to a posterior that has approximately frequentist matching coverage. This is achieved by matching the covariance of the posterior to that of the distribution of true values of the parameters around the maximum likelihood values in repeated trials, under certain assumptions. Using this prior, credible intervals derived from a Bayesian analysis can be interpreted approximately as confidence intervals, containing the truth a certain proportion of the time for repeated trials. Linking frequentist and Bayesian approaches that have previously appeared in the astronomical literature, this offers a consistent and conservative approach for credible intervals quoted on model parameters for problems where the covariance matrix is itself an estimate. 
\end{abstract}

\begin{keywords}
methods: statistical -- methods: data analysis -- cosmology:  observation
\end{keywords}



\section{Introduction}


The problem of fitting a model to multivariate Normal (hereafter referred to as Gaussian) distributed data, where only an approximation to the true data covariance matrix is available, often arises in astrophysics. In a Bayesian sense, the problem can be considered as jointly fitting a model for the data and the covariance matrix, which is a standard one in statistics with a long history. For Gaussian-distributed data, the standard estimate of the covariance matrix is drawn from a Wishart distribution, such as when a covariance matrix is estimated using a limited number of simulations, or when a covariance matrix is constructed from Jackknife samples \citep[e.g.][]{Norberg2009, Friedrich2016}. Examples of cosmological inferences made within this framework include the recent measurements from BOSS and eBOSS \citep{BOSS-DR12-results,eBOSS-DR16-results} as well as the galaxy clustering part of \citet{2021A&A...646A.140H}. For analyses of 2-point statistics in line-of-sight projected data the covariance matrix is often modelled analytically instead of estimating it from simulations \citep[see e.g.\ ][for recent examples]{Krause:2016jvl, 2021A&A...646A.140H, 2021arXiv210513549D}. This is because the 4-point functions constituting those covariances are accurately approximated in a Gaussian model, that is easy to evaluate \citep{2021A&A...646A.129J, 2020arXiv201208568F}. In contrast, analyses of non-standard summary statistics almost exclusively rely on estimated covariances, because analytical covariance models are not easily obtained for them \citep[e.g.][]{Kacprzak2016, Gruen2018, Brouwer2018, Martinet2018, Halder2021}.

There are two common ways to characterise our uncertainty about a model parameter when comparing data and model, which lie at the heart of the difference between Bayesian and frequentist approaches. One can perform a Bayesian analysis using the posterior to define credible intervals, within which a model parameter falls with a particular probability given the prior information and experimental data. One can also define a mechanism to produce frequentist confidence regions, a set proportion of which contain the true parameters in repeated trials. For astrophysical problems we can consider the trials to be experiments performed in parallel universes that are independent and identically distributed realisations of the same data generating process (so the universal constants are considered the same). 
Confidence regions determined, for example, by the distribution of the difference between truth and \corr{the} maximum likelihood solution, will not in general be the same as the credible regions, and it is self-evidently wrong to identify them for asymmetric distributions \citep[see e.g.][]{Loredo12}. That they are not generally the same is evident since credible regions are clearly dependent on the prior, while maximum likelihood estimates are not.  In other words: the fraction of times the credible intervals contain the \emph{true} parameters for repeated analyses (the frequentist coverage probability) is not necessary equal to the posterior probability enclosed within these intervals. The difference has previously been used in astrophysics to search for unrecognized biases during data analysis \citep{ESJL}. 

In this paper, we seek a prior that gives a frequentist matching posterior, so that we can define credible regions that have the property that, for a given parametrisation, the x\% credible regions contain the true parameter values in approximately x\% of repeated trials. This means that we can interpret the mechanism used to define these regions (the Bayesian mechanism) as providing confidence regions with a frequentist coverage probability that matches the Bayesian probability associated with interpreting the same regions as credible regions. This match always holds in the asymptotic limit of infinite data (the Bernstein-von Mises theorem), which includes having a perfect covariance matrix estimate; here our prior ensures the distributions match at the level of equal parameter covariances, for Gaussian linear models and approximately for nonlinear models.

Note that, in general, frequentist matching priors are not a panacea, as they may not perform well in all circumstances, such as in making predictive distributions \citep{SB06}, and they are not invariant to reparametrisation. Note also that the differences between the different priors diminish, as expected, when the number of simulations is large and the posterior is dominated by data.

Before we introduce the problem further and the frequentist matching solution, we introduce the notation adopted: ${\bm x}_0$ are the compressed experimental data of dimension $n_d$ (e.g. a power spectrum), while ${\bm x}_i$ is the simulated data with $1\le i\le n_s$, assumed to be Gaussian distributed around the true model. From the $n_s$ simulations, we construct an unbiased estimate of the covariance matrix $S$,
\begin{equation}  \label{eq:S}
  S = \frac{1}{n_s-1}\sum^{n_s}_{i=1}({\bm x}_i-\bar{\bm x}) 
    ({\bm x}_i-\bar{\bm x})^T\,,
\end{equation}
where $\bar{\bm x}$ is the mean of ${\bm x}_i$ \corr{over all simulations}. \corr{The expectation value of ${\bm x}_0$ is ${\bm \mu}$}, and $\Sigma$ its (unknown) covariance. We only use the simulated data to calculate $S$, and so we consider the data to be $({\bm x}_0,S)$. We will consider fitting a model with $n_\theta$ parameters ${\bm\theta}$, such that our model for the data is ${\bm \mu}({\bm\theta})$, while the covariance matrix used to form the posterior remains of dimension $n_d$. Without loss of generality we shall assume that the expected values of $\mu$ and $\theta$ are zero, such that they can be ignored in our equations and we can, for example, write the covariance for estimates of $\hat{\bm\theta}$ as $\langle\hat{\bm\theta}\hat{\bm\theta}^T\rangle$.

Errors in the covariance matrix used to determine the likelihood have a number of effects on the inferences we make from the data, and particularly the credible intervals quoted in a Bayesian analysis. \citet{hartlap07} was the first to point out in the astronomical literature that, for $S$ calculated using Eq.~\ref{eq:S} and therefore drawn from a Wishart distribution with degrees of freedom $n_s-1$ and scale matrix $\Sigma/(n_s-1)$, $S^{-1}$ is a biased estimator for the inverse covariance matrix $\Sigma^{-1}$, whereas $(hS)^{-1}$ is not, where
\begin{equation}
  h=\frac{n_s-1}{n_s-n_d-2}
\end{equation}
is commonly (by astronomers) called the Hartlap factor \citep[although knowledge of this effect reaches at least as far back as][]{Kaufman}. We discuss the application of the Hartlap factor further in Section~\ref{sec:discussion}.

Taking a frequentist stance, \citet{DS13} and \citet{Taylor2014} showed that the nature of $S$ has a strong effect on the confidence intervals derived based on the distribution of maximum {\em a posteriori} probability (MAP) model parameters (commonly called the best-fit parameters). In fact, we will show later that for the priors and linear models that we consider, the maximum likelihood and \corr{MAP} parameters are the same. So, we could have considered this distribution as the distribution of maximum likelihood solutions. However, as most analyses only work with the posterior, we simply refer to these as the MAP model parameters. \citet{DS13} provided a second order calculation deriving the distribution of MAP model parameters recovered after repeated experiments, averaging over a set of estimated covariance matrices. This derivation is reviewed in Section~\ref{sec:DS}. \citet{Percival14} pointed out that the offset found by \citet{DS13} cannot be applied directly to change credible intervals as the average posterior from a set of repeated experiments itself depends on the distribution of $S$, and they provided a factor by which the credible intervals recovered assuming a Gaussian posterior could be adjusted to match the confidence intervals obtained from the distribution of MAP parameters recovered from mocks. This is discussed further in \corr{Section~\ref{sec:summary}}. 

The Bayesian solution was introduced in the astronomical literature by \cite{SH16} based on the independence Jeffreys prior and marginalising over the unknown covariance matrix. The resulting posterior has multivariate t-distribution form. The derivation follows from Bayes theorem, starting from the joint posterior
\begin{equation}  \label{eq:joint_post}
  f({\bm \mu},\Sigma|{\bm x}_0,S) \propto f({\bm x}_0,S|{\bm
    \mu},\Sigma)f({\bm \mu},\Sigma)\,,
\end{equation}
where $f({\bm \mu},\Sigma)$ is the prior, and $f({\bm x}_0,S|{\bm\mu},\Sigma)$ the likelihood. Because of the independence of ${\bm x}_0$ and $S$, the likelihood can be written
\begin{equation}
  f({\bm x}_0,S|{\bm\mu},\Sigma) = f({\bm x}_0|{\bm\mu},\Sigma)f(S|\Sigma)\,.
\end{equation}
To make model inferences, we wish to know the distribution of the data-generating mechanism (or its parameters) given the data and $S$, which we can calculate by marginalising over the true covariance:
\begin{equation}
  f({\bm \mu}|{\bm x}_0,S)=\int d\Sigma\ f({\bm \mu},\Sigma|{\bm x}_0,S)\,.
\end{equation}

The key question in a Bayesian analysis performed under these conditions is the form for the prior proposed for $\mu$ and the covariance matrix. \citet{SB06} listed a number of options for prior choices, including the Jeffreys prior,
\begin{equation}
  f({\bm \mu},\Sigma)\propto|\Sigma|^{-\frac{n_d+2}{2}}\,,
\end{equation}
and independence Jeffreys prior (adopted by \citealt{SH16}),
\begin{equation}
  f({\bm \mu},\Sigma)\propto|\Sigma|^{-\frac{n_d+1}{2}}\,.
  \label{eq:IJ}
\end{equation}
\corr{\citet{GC63} consider a range of priors 
\begin{equation}
  f({\bm \mu},\Sigma)\propto|\Sigma|^{-v}\,,
\end{equation}
where $v$ is an integer with $v\le n_s$.} Various other potential priors have also been introduced (e.g. Haar prior, right-Haar prior, left-Haar prior, \citealt{CE90} reference prior) with more complicated forms. Each has advocates and interesting properties in various situations.

The prior that we introduce is a member of the class of frequentist matching priors \citep{Lindley58,Welch63,Reid03}, designed to match a posterior to frequentist expectations. A discussion of such priors is given in \cite{Ghosh11}. Priors that match posterior predictive probabilities with the corresponding frequentist probabilities are attractive when constructing credible / confidence intervals. In general, matching priors can be constructed only for particular models and matching is determined by the order of approximation to the integrated probability. The selection of a matching prior is usually accompanied by a discussion of the degree of matching, with various definitions of matching available (e.g. \citealt{Reid03}). Although matching is usually considered between cumulative probabilities, we match on the expected model parameter covariance. This second moment is commonly used as the basis for model parameter confidence intervals in physics, and can be broadly interpreted as fixing the multi-dimensional ``width" of a distribution. 

Matching priors are candidates for non-informative priors in Bayesian inference, \corr{in that it is often assumed (explicitly or not) that the frequentist-style determination of confidence intervals incorporates no information from a prior.} Really, there is simply no such thing as a non-informative prior. The frequentist philosophy is different from the Bayesian approach and provides different guarantees across notionally repeated experiments. However, given that the concept of "errors" is often interpreted according to the frequentist philosophy, we think there is merit in making the widths of the errors 
consistent.

Matching priors (and frequentist analyses) violate the Likelihood Principle by using priors that vary with the sampling distribution of the experiment to be performed and the dimension of the model parameter space onto which the data distribution is projected. However, in general they only rely on the performance characteristics of that distribution under repeated sampling, as a way to “break the tie” among a choice of prior distributions, in order to draw an inference. Thus, while there is debate about their validity and usage, it is clear that there are situations where they are useful. 

In this paper, we argue that the analyses presented in \citet{hartlap07} and \citet{DS13} provide a method for calculating frequentist based confidence intervals for model parameters, and we show that these can be matched to credible intervals obtained from a Bayesian analysis as advocated by \citet{SH16}. A similar calculation was performed by \citet{Percival14} but we now use the methodology and resulting form for the posterior adopted by \citet{SH16}, albeit using a different prior. 
This demonstrates how these different methods are related and the different assumptions being (sometimes implicitly) made when adopting one of these \corr{procedures} for determining and quoting the coverage probability associated with an interval. The frequentist matched credible intervals are larger than those from Bayesian analyses with previously used priors, and hence this matching can also be considered conservative for inferences made from experiments. 

The layout of our paper is as follows: Section~\ref{sec:bayesian} introduces the Bayesian problem that we want to solve, and considers how the posterior depends on the prior chosen, extending the \citet{SH16} approach to more general priors. Section~\ref{sec:confidence-intervals} considers probabilities under the posterior and relates them to the distribution of the truth after repeated trials, allowing us to define a frequentist matching prior in Section~\ref{sec:matching}. Section~\ref{sec:MC_test} demonstrates this approach using the simple problem of fitting a mean to correlated data, using both analytic derivations and Monte Carlo simulations. We apply our approach to a realistic cosmological analysis in Section~\ref{sec:DES_like}, fitting mock tomographic cosmic shear data vector including auto- and cross-correlations matching that expected from the 5-year data of the Dark Energy Survey, demonstrating that this works well in a practical test, providing Bayesian credible intervals on model parameters that \corr{match} the expected frequentist confidence intervals. We summarise our proposed method in Section~\ref{sec:summary}, and conclude in Section~\ref{sec:discussion}.

\section{Choice of prior to use in a model fit}  \label{sec:bayesian}

In this section we consider a full Bayesian analysis of the problem, considering different choices for the prior.

\subsection{Posterior with an independence Jeffreys prior}  \label{sec:jeff-prior}

The uninformative nature of the independence Jeffreys prior in general was introduced at the very start of Bayesian statistics \citep{Jeffreys1939} and is discussed in this specific situation in \citet{SB06}. It assumes for Gaussian data a uniform prior for the means, and a Jeffreys prior for the covariance matrix with means given \citep{BergerSun08}. The derivation of the posterior using this choice of prior, and application to astronomical situations was presented in \citet{SH16}.



We assume the independence Jeffreys joint prior on the \corr{expectation value of the} data and its covariance matrix given by Eq.~\ref{eq:IJ}.
To calculate the required posterior using Eq.~\ref{eq:joint_post}, we first note that $S$ follows a Wishart distribution, $f_W$, and we can write
\begin{eqnarray}
  f(\Sigma|S) &\propto& 
    f_W(S|\Sigma/\corr{(n_s-1)},n_s-1)f({\bm \mu},\Sigma)\,,\\
  &\propto&
            |\Sigma|^{-\frac{n_s+n_d}{2}}\exp\left[-\frac{n_s-1}{2}Tr(\Sigma^{-1}S)\right]\,,\\
  &\propto& f_{W^{-1}}(\Sigma|(n_s-1)S,n_s-1)\,,
\end{eqnarray}
which shows how, with this prior, the posterior for $\Sigma$ has an inverse Wishart distribution, $f_{W^{-1}}$. \corr{The definitions of the multivariate distributions used in our work are included in Appendix~\ref{sec:mult-dists}}.

We now multiply by the Gaussian likelihood $f_N({\bm x}_0|{\bm \mu},\Sigma)$, which is simplest to consider in the form given in Appendix~\ref{sec:mult-dists}, and integrate over $\Sigma$ to find that
\begin{equation}
  f({\bm \mu}|{\bm x}_0,S)\propto\int d\Sigma\ 
  |\Sigma|^{-\frac{n_s+n_d+1}{2}}\exp\left[-\frac{1}{2}Tr(\Sigma^{-1}Q)\right]\,,
\end{equation}
where
\begin{equation}  \label{eq:Q}
  Q = (n_s-1)S+({\bm x}_0-{\bm \mu})({\bm x}_0-{\bm \mu})^T\,.
\end{equation}
This is an integral over the unnormalised inverse Wishart distribution (with parameter $n_s$), so we can read off the result from the normalisation constant in Eq.~\ref{eq:iwishart}.
\begin{equation}
  f({\bm \mu}|{\bm x}_0,S)\propto |Q|^{-\frac{n_s}{2}}\,.
\end{equation}
Comparing with the form of the multivariate t-distribution in \corr{Eq.~\ref{eq:tdist}}, we see that
\begin{equation}
    f({\bm \mu}|{\bm x}_0,S)=f_{t,n_s-n_d}\left(\corr{\bm\mu}\left|{\bm x}_0,\frac{n_s-1}{n_s-n_d}S\right.\right)\,,
\end{equation}
which has mean ${\bm x}_0$ and covariance 
\begin{equation}  \label{eq:var_tdist_ns}
  \langle({\bm\mu}-{\bm x}_0)({\bm\mu}-{\bm x}_0)^T\rangle
    =\frac{n_s-1}{n_s-n_d-2}S
    =hS\,.
\end{equation}

The use of the multivariate t-distribution as a replacement for the Gaussian assumption is often advocated on the grounds of robustness to outliers \citep{Lange89}, with the parameter $\nu$, which in our context is $n_s-n_d$ used as a robustness tuning factor. In this section we have shown how it also arises when the covariance matrix is itself a random variable. It is also interesting to see that, with an independence Jeffreys prior, the Hartlap factor emerges in the recovered covariance, which could be considered natural given that using this prior brings in no further information on the posterior, and the inclusion of the Hartlap factor in some sense unbiases the posterior covariance. However, inferences made from the posterior about the covariance on model parameters are biased by the inclusion of this factor - while it unbiases the posterior against repeated trials of $S$, inferences about model parameter covariances made from the posterior are biased - and so it is not clear that this is what we actually want (see Section~\ref{sec:discussion} for further discussion of this). We also note that a Gaussian posterior with a Hartlap correction yields a posterior covariance that agrees with that derived here, but has tail probabilities that are lower than the t-distribution, and may be in considerable error when datasets in tension are discussed and compared (see Appendix~\ref{sec:intervals}). 

In the next section we see that the multivariate t-distribution form for the posterior follows from any prior that is a power-law in $|\Sigma|$, and that the exponent of the power-law affects the recovered credible intervals. 

\subsection{Posterior with a general power-law prior}

Let us now consider a more general joint prior on the mean and covariance matrix
\begin{equation}  \label{eq:mprior}
  f({\bm \mu},\Sigma)\propto|\Sigma|^{-\frac{m-n_s+n_d+1}{2}}\,.
\end{equation}
The independence Jeffreys prior of \cite{SH16} corresponds to $m=n_s$.  Both priors are uniform in the mean, which makes sense for a location parameter. The exact linear form for the exponent is chosen to simplify the downstream analysis, but is not important. It changes our conditional likelihood
\begin{equation}
  f(\Sigma|S)
  \propto |\Sigma|^{-\frac{m+n_d}{2}}
    \exp\left[-\frac{n_s-1}{2}Tr(\Sigma^{-1}S)\right]\,,
\end{equation}
and we now have that
\begin{equation}
  f({\bm \mu}|{\bm x}_0,S)\propto\int d\Sigma
  |\Sigma|^{-\frac{m+n_d+1}{2}}\exp\left[-\frac{1}{2}Tr(\Sigma^{-1}Q)\right]\,,
\end{equation}
where $Q$ is given by Eq.~\ref{eq:Q}. The form of this equation still matches that of an unnormalised inverse Wishart distribution, but with different parameters, so we now have
\begin{equation}
  f({\bm \mu}|{\bm x}_0,S)\propto |Q|^{-\frac{m}{2}}\,.
\end{equation}
Following through the derivation,
\begin{equation}  \label{eq:tdist-gen}
      f({\bm \mu}|{\bm x}_0,S)=f_{t,m-n_d}\left(\corr{\bm\mu}\left|{\bm x}_0,\frac{n_s-1}{m-n_d}S\right.\right)\,.
\end{equation}
From the known properties of the multivariate t-distribution, this has mean ${\bm x}_0$ and covariance
\begin{equation}  \label{eq:var_tdist_m}
  \langle({\bm\mu}-{\bm x}_0)({\bm\mu}-{\bm x}_0)^T\rangle
    =\frac{n_s-1}{m-n_d-2}S\,.
\end{equation}
As expected, setting $m=n_s$ gets us back to Eq.~\ref{eq:var_tdist_ns}, and an expected covariance of $hS$. The covariance recovered from the distribution is directly related to the prior through $m$ - as is natural in a Bayesian analysis.

\section{Model parameter covariances from posteriors and from the parameter distribution}  \label{sec:confidence-intervals}

We now consider different methods for characterising our uncertainty about model parameters by comparing the model parameter covariances calculated using different assumptions. 

Given a set of data $({\bm x}_0,S)$ and a prior parameterised by $m$, we first determine the Fisher matrix (Section~\ref{sec:Fish}) and then consider the model parameter covariance derived by computing probabilities under the posterior (Section~\ref{sec:Var}). In order to construct a matching prior, for probabilities estimated using the Fisher matrix and probabilities calculated under the posterior, we need to determine the frequentist coverage probability that can be associated with the derived credible intervals. Formally, the coverage probability is a property of the procedure for constructing frequentist confidence intervals, and gives the proportion of repeated trials for which the interval contains the true value of interest. As we want to be able to interpret x\% credible intervals as x\% confidence intervals, we need to calculate the average size of the credible intervals of fixed probability over repeated trials. Finding the prior for which this is equal to the probability of finding the truth within each interval after repeated trials would then mean that we could interpret Bayesian credible intervals containing a particular probability with the same coverage probability. For simplicity, we work with the covariance rather than the intervals directly and hence we wish to know the average model parameter covariance recovered from the Fisher matrix or the posterior over repeated trials. For this, the multivariate t-distribution posterior has some differences from the expectation for a Gaussian posterior because the covariance of the posterior around the MAP model parameters depends on ${\bm x}_0$ in addition to $S$. Consequently, the distribution assumed for the data is important as we demonstrate by contrasting results assuming the data is drawn from a t-distribution, or from a Gaussian as is correct for our problem. The dependence of the model parameter covariance on ${\bm x}_0$ also affects data compression as we show in Appendix~\ref{sec:Compress}.

We contrast the covariance estimated by integrating under the posterior with that calculated for the distribution of MAP solutions given the truth in Section~\ref{sec:DS}, formally showing that, for our problem, they are very different for most choices of prior. In Section~\ref{sec:matching} we present the prior that matches these results.

\subsection{Using the Fisher matrix}  \label{sec:Fish}

The Fisher information matrix (or simply the Fisher matrix), defined as
\begin{equation}
    F({\bm\theta})_{\alpha\beta}=E\left[
    \left(\frac{\partial}{\partial\theta_\alpha}
      \log f({\bm x}_0|{\bm\theta})\right)
    \left(\frac{\partial}{\partial\theta_\beta}
      \log f({\bm x}_0|{\bm\theta})\right)\right]\,,
\end{equation}
is a function of the likelihood. In Bayesian inference, the Bernstein-von Mises theorem provides the basis for using the Fisher matrix to provide confidence statements on parametric models, and the Cram\'er–Rao theorem shows that it forms a lower bound for the covariance of unbiased estimators of $\bm\theta$. In our case, we work from the posterior, as given in Eq.~\ref{eq:tdist-gen}, and convert this to a likelihood assuming a uniform prior (albeit possibly improper) on the model parameters. Thus, in this section, we are not calculating the Fisher matrix from the true likelihood of the data (remember that ${\bm x}_0$ are drawn from a Gaussian distribution with covariance $\Sigma$), but instead we use the Fisher matrix to estimate the expected information given the form of the posterior assumed.

We start by assuming that, around the peak of the posterior, we can define a patch of parameter space for which we can apply Bayes theorem to Eq.~\ref{eq:tdist-gen} with a uniform prior on ${\bm\mu}$. For this patch 
the likelihood for ${\bm x}_0$ is 
\begin{equation}  \label{eq:tdist-x0}
      f({\bm x}_0|{\bm \mu},S) = f_{t,m-n_d}\left(\corr{\bm x}_0\left|{\bm\mu},\frac{n_s-1}{m-n_d}S\right.\right)\,.
\end{equation}
The Fisher information matrix for the multivariate t-distribution with degrees of freedom $\nu$ and covariance $\Sigma$ \citep{Lange89,SH17} is
\begin{equation}  \label{eq:Ft}
  F_t = \frac{\nu(\nu+n_d)}{(\nu-2)(\nu+n_d+2)}
    \frac{\partial {\bm\mu}}{\partial {\bm\theta}}^T \Sigma^{-1}
    \frac{\partial {\bm\mu}}{\partial {\bm\theta}}\,.
\end{equation}
We see an extra term compared with the true Fisher Information matrix if the covariance matrix were known:
\begin{equation}  \label{eq:Ftrue}
  F_\Sigma =
  {\frac{\partial {\bm\mu}}{\partial {\bm\theta}}}^T \Sigma^{-1}
  {\frac{\partial {\bm\mu}}{\partial {\bm\theta}}}\,.
\end{equation}
For completeness, the Gaussian Fisher Information matrix with covariance matrix $S$ is 
\begin{equation}  \label{eq:Fsdist}
  F_S =
  {\frac{\partial {\bm\mu}}{\partial {\bm\theta}}}^T S^{-1}
  {\frac{\partial {\bm\mu}}{\partial {\bm\theta}}}\,.
\end{equation}
For the likelihood of Eq.~\ref{eq:tdist-x0}, we have $\nu=m-n_d$ degrees of freedom and a covariance $(n_s-1)S/(m-n_d-2)$, so we have
\begin{equation}  \label{eq:Ftdist}
  F_t = \frac{m(m-n_d)}{(m+2)(n_s-1)}
  F_S\,.
\end{equation}
This is the t-distribution Fisher matrix given the approximate scale matrix $S$.

As discussed at the start this section, we also want to determine the average credible interval that would be recovered given a set of realisations of $S$ drawn from a Wishart distribution (i.e. by observers in parallel universes). To calculate this, we note that a property of the Wishart distribution is that for
\begin{equation}
  f(S|\Sigma)=f_W(S|\Sigma/(n_s-1),n_s-1)\,,
\end{equation}
and $M$ a $n_\theta\times n_d$ matrix, then 
\begin{displaymath}
\corr{f(\,(MS^{-1}M^T)^{-1}|\Sigma) =}
\end{displaymath}
\begin{equation}  \label{eq:MSinvM}
  \corr{f_W\left((MS^{-1}M^T)^{-1}\left|
    \frac{(M\Sigma^{-1}M^T)^{-1}}{n_s-1},
        n_s-n_d+n_\theta-1\right.\right)}\,, 
\end{equation}
(see theorem 3.2.11 of \citealt{Muirhead1982}). Thus, from Eq.~\ref{eq:Fsdist}, \corr{and using the mean of the Wishart distribution,} we have that
\begin{equation}  \label{eq:invFish-Gauss}
  \langle F_S^{-1}\rangle_S =\frac{n_s-n_d+n_\theta-1}{n_s-1}F_\Sigma^{-1}\,.
\end{equation}
This equation can also be approximated by writing $(hS)^{-1}$ as a perturbation around $\Sigma^{-1}$ and considering the second order terms, as discussed in Appendix~\ref{sec:pert}, and used in \citet{Percival14}. 

For the t-distribution Fisher matrix, from Eq.~\ref{eq:Ftdist}, we have that
\begin{equation}  \label{eq:var_fish}
  \langle F_t^{-1}\rangle_S 
    = \frac{(m+2)(n_s-n_d+n_\theta-1)}{m(m-n_d)}
      F_\Sigma^{-1}\,.
\end{equation}
This shows that the error in the covariance matrix has an additional effect on the average model parameter credible intervals derived from a set of realisations of the scale matrix. 

\subsection{Computing probabilities under the posterior}  \label{sec:Var}

We now consider credible intervals derived by computing probabilities under the posterior, based on the 2nd moment of the distribution. While the Fisher matrix gives the form of the likelihood around the expected value, calculating probabilities under the posterior is the more common approach used for model parameter credible interval determination. We consider the case where we have a linear model with ${\bm\mu}=E{\bm\theta}$, for some generally non-square matrix $E$. \corr{Using Eq.~\ref{eq:tdist-gen} the posterior can be written}
\begin{equation}  \label{eq:tdist-post-linear1}
  f({\bm\theta}|{\bm x}_0,S) \propto
  \left[ 1 + \frac{1}{n_s-1}\left({\bm x}_0-E{\bm\theta}\right)^T 
    S^{-1}\left({\bm x}_0-E{\bm\theta}\right)\right]^{-\frac{m}{2}}\,.
\end{equation}
This can be manipulated to describe the posterior as a distribution around the MAP estimate. For a simple example of this for a Gaussian posterior, and a \corr{single-parameter} model - fitting the mean to data - see Appendix~\ref{sec:mean-gauss}. The same derivation can be seen in Appendix~\ref{sec:mean-tdist} for the case of fitting the mean using a t-distribution posterior. Keeping to a more general linear model, expanding the distribution, we have
\begin{equation}  \label{eq:tdist-post-linear2}
  f({\bm\theta}|{\bm x}_0,S) \propto
  \left[ 1 + \frac{
    {\bm x}_0^TS^{-1}{\bm x}_0 -
    2{\bm\theta}^TE^TS^{-1}{\bm x}_0 +
    {\bm\theta}^TE^TS^{-1}E{\bm\theta}
  }{n_s-1}\right] ^{-\frac{m}{2}}\,,
\end{equation}
using the symmetry of $S^{-1}$ to simplify the cross terms. Setting $F_S=E^TS^{-1}E$ and ${\bm g}=E^TS^{-1}{\bm x}_0$ gives
\begin{equation} \label{eq:tdist-post-linear3}
  f({\bm\theta}|{\bm x}_0,S) \propto
  \left[ 1 + \frac{
  {\bm x}_0^TS^{-1}{\bm x}_0-{\bm g}^TF_S^{-1}{\bm g} +
  ({\bm\theta}-F_S^{-1}{\bm g})^TF_S({\bm\theta}-F_S^{-1}{\bm g})
  }{n_s-1}\right] ^{-\frac{m}{2}}\,.
\end{equation}
To finish the derivation, we need to complete the square, noting that if we now define
\begin{equation} \label{eq:ylin}
  {\bm y} = ({\bm\theta}-F_S^{-1}{\bm g})
  \left(\frac{n_s-1}{m-n_\theta}\right)^\corr{-\frac{1}{2}}
  \left[1 + \frac{{\bm x}_0^TS^{-1}{\bm x}_0-{\bm g}^TF_S^{-1}{\bm g}}{n_s-1}\right]^\corr{-\frac{1}{2}}\,,
\end{equation}
then the posterior reduces to the simple form
\begin{equation}  \label{eq:tdist-post-y}
  f({\bm\theta}|{\bm x}_0,S) \propto
  \left[ 1 + \frac{{\bm y}^TF_S^{-1}{\bm y}}{m-n_\theta}\right]^{-\frac{m}{2}}\,.
\end{equation}
This shows that ${\bm y}$ is distributed with a multivariate t-distribution with $m-n_\theta$ degrees of freedom, such that the mean $\langle {\bm y}\rangle={\bm 0}$, and covariance $\langle {\bm y}{\bm y}^T\rangle=(m-n_\theta)\corr{F_S^{-1}}/(m-n_\theta-2)$.

We can write ${\bm\theta}$ in the form ${\bm\theta}=a{\bm y}+b$, which has the property that $\langle{\bm\theta}\rangle=\corr{a}\langle{\bm y}\rangle+b$, and $\langle({\bm\theta}-\hat{\bm\theta})({\bm\theta}-\hat{\bm\theta})^T\rangle=a^2\langle {\bm y}{\bm y}^T\rangle$. From this, we see that the distribution of ${\bm\theta}$ has mean $\hat{\bm\theta}=\langle{\bm\theta}\rangle=F_S^{-1}{\bm g}$. The covariance of ${\bm\theta}$ around this for any value of ${\bm x}_0$ and $S$ is
\begin{displaymath}
\langle({\bm\theta}-\hat{\bm\theta})({\bm\theta}-\hat{\bm\theta})^T\rangle =
\end{displaymath}
\begin{equation} \label{eq:var_post_givenxandS}
    \hspace{1cm}
    \frac{n_s-1}{m-n_\theta-2} F_S^{-1}
    \left[1 + \frac{{\bm x}_0^TS^{-1}{\bm x}_0-{\bm g}^TF_S^{-1}{\bm g}}{n_s-1}\right]\,.
\end{equation}
For a linear model, this expression can be used instead of integrating under the posterior for any realisation of the data $({\bm x}_0,S)$. Crucially, unlike the equivalent calculation for the Gaussian distribution (see Appendix~\ref{sec:mean-gauss} for this calculation in the special case of fitting the mean to data), the model parameter covariance depends on the value of ${\bm x}_0$. Thus, the size of the credible intervals we derive from our fit will change if we change the data. 

We now consider the model parameter covariance recovered by integrating under the posterior, averaged over a set of values of ${\bm x}_0$ and $S$. We start by considering ${\bm x}_0$ distributed according to the t-distribution, and a Wishart distributed $S$. However, while we adopt a posterior that has multivariate t-distribution form, the data itself are actually Gaussian distributed with covariance $\Sigma$, and so we consider this case afterwards.

\subsubsection{Data distributed according to the t-distribution}  \label{sec:var_post_tdist}

We can now calculate the expected covariance recovered for the model parameters, averaging over multiple realisations of the data $({\bm x}_0,S)$. We start by assuming that the same covariance matrix approximation $S$ is used for all realisations. In this case, $F_S^{-1}$ is fixed, and we need to replace the terms ${\bm x}_0^TS^{-1}{\bm x}_0$ and ${\bm g}^TF_S^{-1}{\bm g}$ by the relevant expected values. \corr{To calculate these, we make use of the fact that we have set up the problem such that $\langle {\bm x}_0\rangle$ is the zero vector, and make use of the identity ${\bm x}_0^TS^{-1}{\bm x}_0=Tr(S^{-1}{\bm x}_0{\bm x}_0^T)$. We find that,} for a set of data drawn from a multivariate t-distribution as in Eq.~\ref{eq:tdist-x0}, we have 
\begin{eqnarray}
  \langle{\bm x}_0^TS^{-1}{\bm x}_0\rangle_x&=&\frac{n_s-1}{m-n_d-2}n_d\,,\\
  \langle{\bm g}^TF_S^{-1}{\bm g}\rangle_x&=&\frac{n_s-1}{m-n_d-2}n_\theta\,.
\end{eqnarray}
Putting these values in to Eq.~\ref{eq:var_post_givenxandS}, the covariance for ${\bm\theta}$ reduces to
\begin{equation}  \label{eq:cov-theta-tdist-mS}
    \langle({\bm\theta}-\hat{\bm\theta})({\bm\theta}-\hat{\bm\theta})^T\rangle_{x}=\frac{n_s-1}{m-n_d-2}F_S^{-1}\,.
\end{equation}
The expectation over multiple $S$ matrices drawn from a Wishart distribution can easily be calculated using Eq.~\ref{eq:invFish-Gauss},
\begin{equation}  \label{eq:cov-theta-tdist-m}
    \langle({\bm\theta}-\hat{\bm\theta})({\bm\theta}-\hat{\bm\theta})^T\rangle_{x,S}=\frac{n_s-n_d+n_\theta-1}{m-n_d-2}F_\Sigma^{-1}\,.
\end{equation}

\subsubsection{Gaussian distributed data}  \label{sec:var_post_gauss}

For a set of data drawn from a Gaussian distribution with covariance $\Sigma$, we have
\begin{eqnarray}  \label{eq:expected_xsq}
  \langle{\bm x}_0^TS^{-1}{\bm x}_0\rangle_x &=& Tr[S^{-1}\Sigma]\,,\\
  \langle{\bm g}^TF_S^{-1}{\bm g}\rangle_x &=& Tr[F_S^{-1}E^TS^{-1}\Sigma S^{-1}E]\,.
\end{eqnarray}

To go one step further and consider the expected model parameter covariance allowing for multiple $S$ matrices drawn from a Wishart distribution, we now need to find expressions for the expectation of all of the terms in Eq.~\ref{eq:var_post_givenxandS}. We have Eq.~\ref{eq:invFish-Gauss} for $\langle F_S^{-1}\rangle_S$, and
\begin{eqnarray}
  \langle F_S^{-1}Tr[S^{-1}\Sigma]\rangle_S \simeq   [n_d+B(n_d(n_\theta+1)-2)]F_\Sigma^{-1}\,,\label{eq:pert-mid}\\
  \langle F_S^{-1}Tr[F_S^{-1}E^TS^{-1}\Sigma S^{-1}E]\rangle_S \simeq
  [n_\theta+B(n_\theta(n_d+1)-2)]F_\Sigma^{-1}
  \label{eq:pert-end}\,,
\end{eqnarray}
where $B$ is given in Eq.~\ref{eq:AB}. To get these \corr{expressions}, we have used the perturbative expressions as described in Appendix~\ref{sec:pert}. 

The end result is that we should expect the average model parameter covariance recovered integrating under the posterior after repeated trials where the data is drawn from a Gaussian distribution with true covariance $\Sigma$, and $S$ is drawn from a Wishart distribution to be
\begin{equation} \label{eq:var_post_overxandS}
  \langle({\bm\theta}-\hat{\bm\theta})({\bm\theta}-\hat{\bm\theta})^T\rangle_{x,S}
    \simeq \frac{n_s-1+B(n_d-n_\theta)}{m-n_\theta-2}F_\Sigma^{-1}\,,
\end{equation}
to second order. The difference between this expression and that of Eq.~\ref{eq:cov-theta-tdist-m} shows the importance of the distribution of ${\bm x}_0$ in calculating the average model parameter covariance recovered. The situation with Gaussian distributed data matches the setup of our problem: that of considering observers in multiple universes.

\subsection{The distribution of the difference between MAP estimate and the truth} \label{sec:DS}

We now contrast these estimates of the model parameter covariance against the distribution of recovered maximum {\em a posteriori} model parameter values recovered from reruns of the experiment being performed. A linear model is assumed, so we have the symmetry that the distribution of MAP solutions about the truth is the same as the distribution of the truth around a particular MAP solution (when the truth is sampled from a uniform prior). \corr{By comparing the results in Section~\ref{sec:Var} to those from a Gaussian posterior,} we see that the MAP estimate for the model parameters is the same whether using a Gaussian or t-distribution posterior and so we do not need to distinguish between these choices.

We therefore start assuming a Gaussian posterior distribution as in \citet{DS13}. As discussed in Section~\ref{sec:Var}, the MAP estimate for a linear model can be written 
\begin{equation}  \label{eq:thetahat}
  \hat{\bm \theta}=F_S^{-1}{\bm g}=F_S^{-1}E^TS^{-1}{\bm x}_0\,,
\end{equation}
which can also be recovered as the first order solution for more general models by Taylor expanding \corr{the} posterior around the MAP estimates of the model parameters. Here we have assumed, without loss of generality, that the true values are $\hat{\bm\theta}={\bm 0}$.

We can now obtain an estimate of the scatter on model parameters provided by different experiments, where we consider different ${\bm x}_0$ drawn from a Gaussian distribution, and $S$ from a Wishart distribution given the true model $\langle\hat{\bm\theta}^T\hat{\bm\theta}\rangle_{x,S}$. To do this, we use the fact that $\langle {\bm x}_0{\bm x}_0^T\rangle_x=\Sigma$, so that
\begin{equation}
  \langle\hat{\bm\theta}\hat{\bm\theta}^\corr{T}\rangle_{x}
    =\langle F_S^{-1}E^TS^{-1}\Sigma S^{-1}EF_S^{-1}\rangle\,.
\end{equation}
This can be solved to second order, using the expression in \corr{Eq.~\ref{eq:cov-wish}}, considering an expansion of $(hS)^{-1}$ around $\Sigma^{-1}$. As described in Appendix~\ref{sec:pert}, the second order solution is 
\begin{equation}  \label{eq:var_dist}
  \langle\hat{\bm\theta}\hat{\bm\theta}^\corr{T}\rangle_{x,S}
    \simeq\left[1+B(n_d-n_\theta)\right]F_\Sigma^{-1}\,,
\end{equation}
which is the distribution of MAP estimates made from a set of simulations that is independent of those used to estimate the covariance matrix $S$. This was the primary result of \citet{DS13}. Because we assume a linear model, this model parameter covariance is also that of the distribution of the truth around the MAP solution, assuming a uniform prior on the model parameters. It is therefore the covariance of the distribution from which frequentist confidence intervals on model parameters are derived.

\begin{figure*}
  \centering
    \resizebox{0.45\textwidth}{!}{\includegraphics{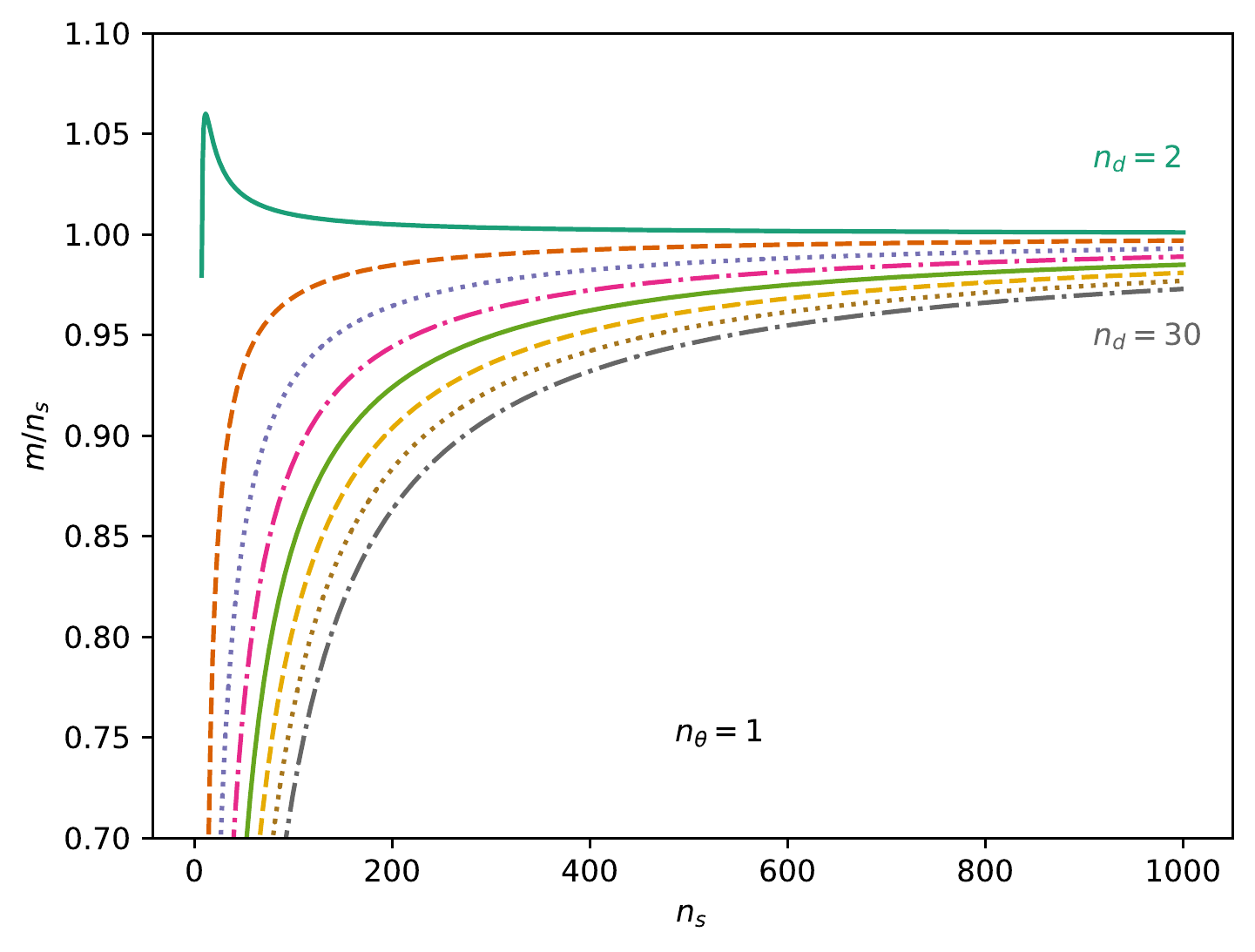}}
  \quad
    \resizebox{0.45\textwidth}{!}{\includegraphics{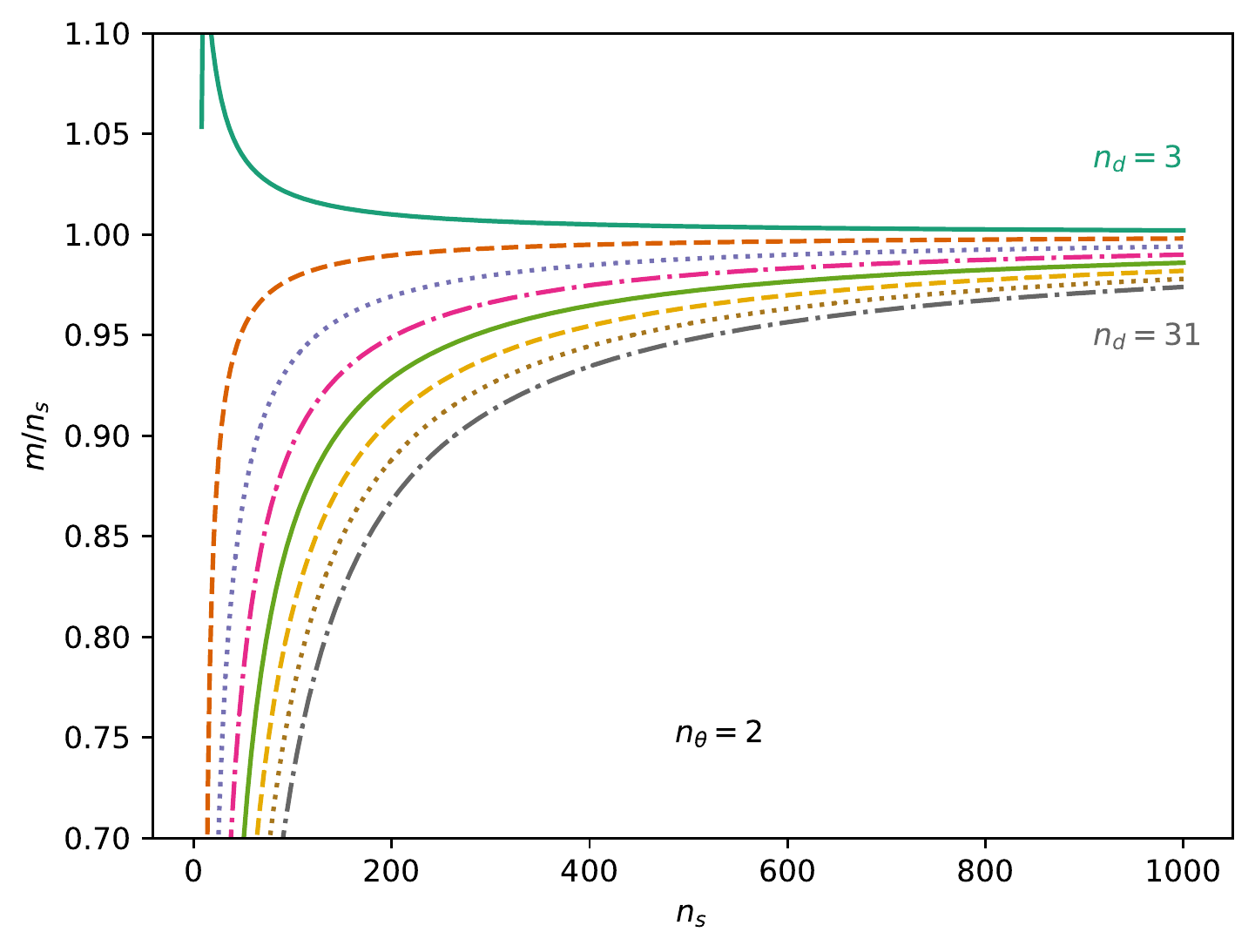}}
\newline 
    \resizebox{0.45\textwidth}{!}{\includegraphics{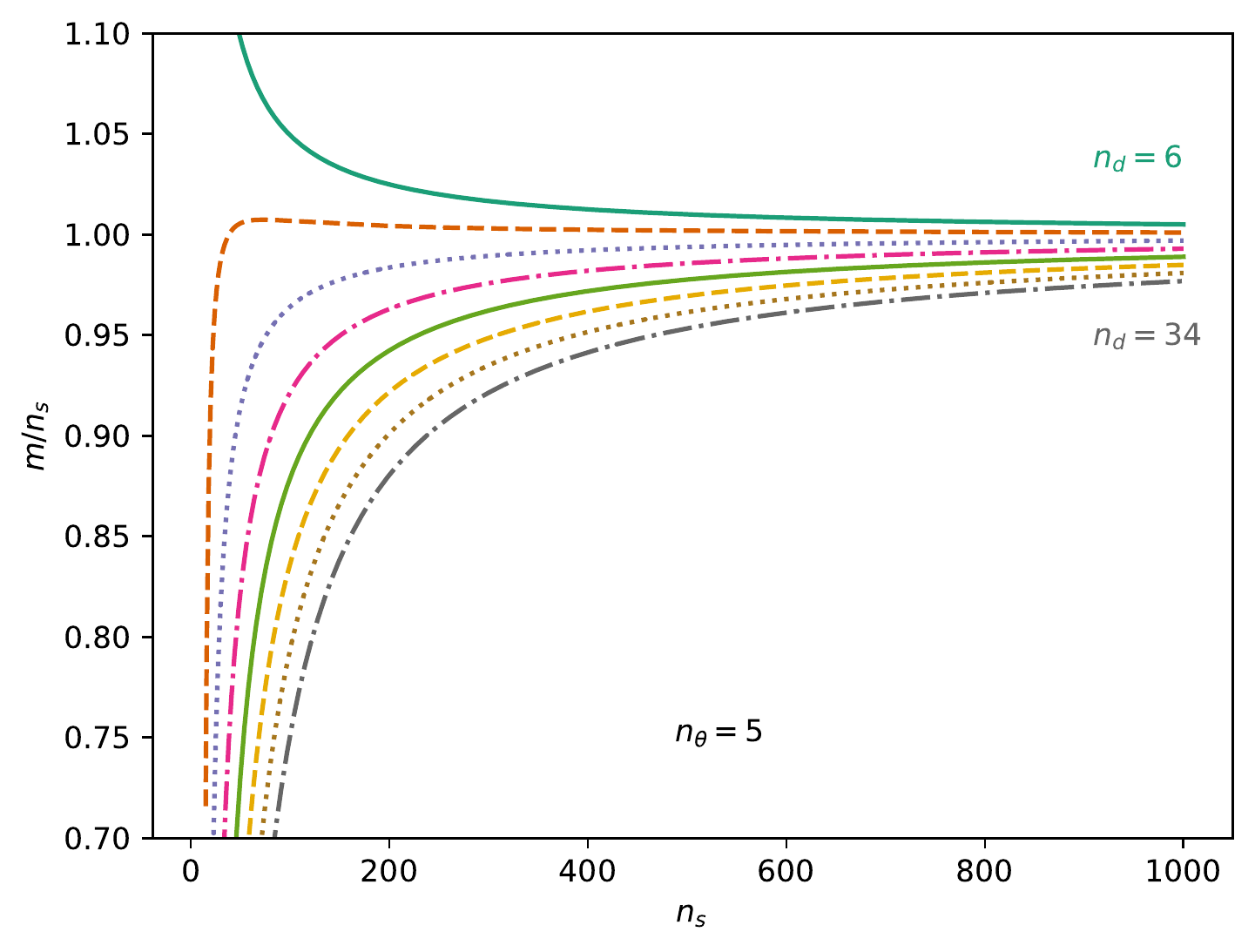}}
  \quad
    \resizebox{0.45\textwidth}{!}{\includegraphics{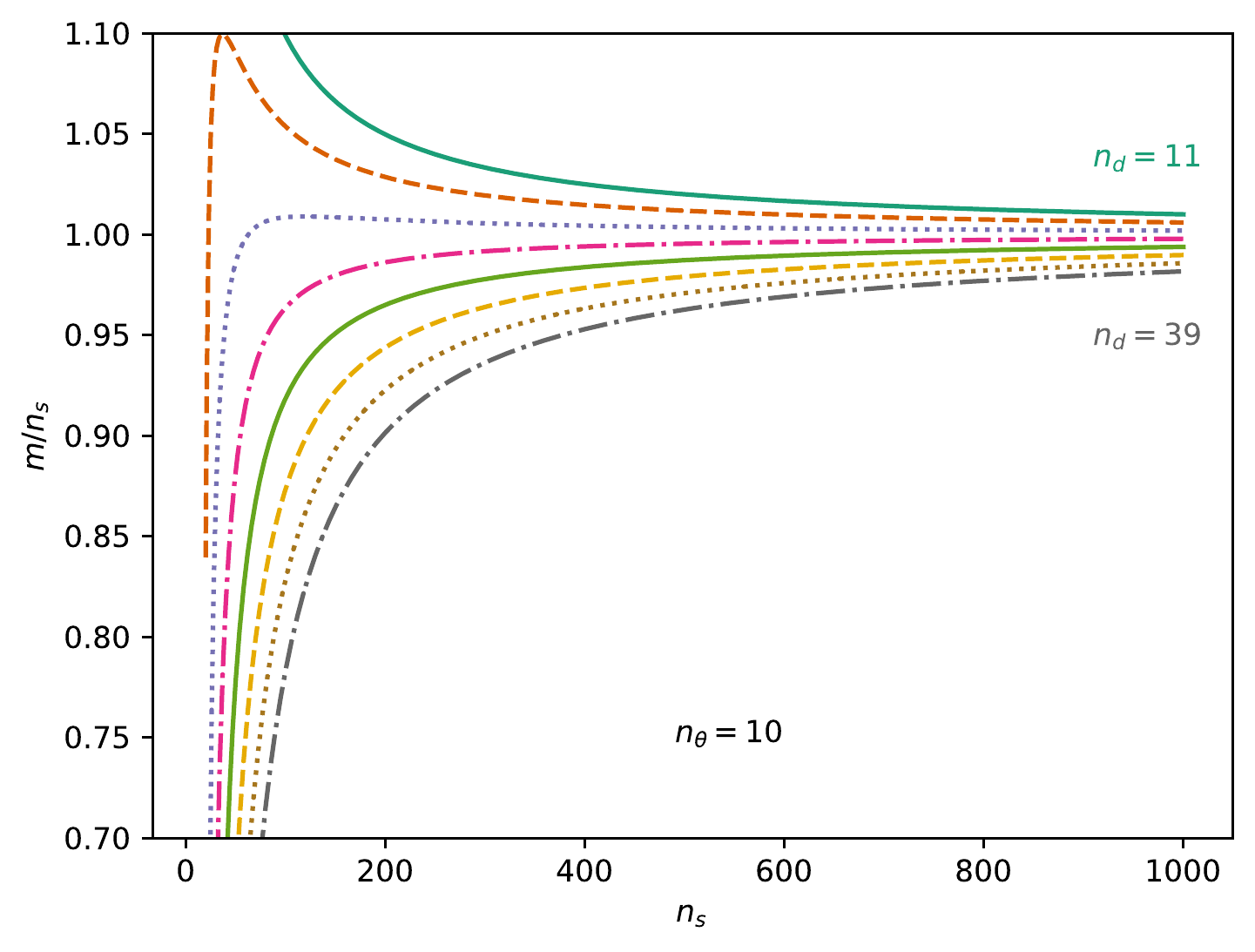}}
  \caption{Variation of the power law exponent $m_\textrm{match}$ required for a prior that, on average over repeated trials gives a posterior with covariance that matches that expected for the distribution of MAP values (solid lines as given in Eq.~\ref{eq:M}). We show how $m_\textrm{match}/n_s$ varies with $n_s$ (x-axis), $n_d$ (different lines) and $n_\theta$ (different panels). \label{fig:M}}
\end{figure*}

\section{frequentist matching prior}  \label{sec:matching}

We now consider how to derive a matching prior that will allow the average model parameter covariance derived from the Bayesian analysis described above to match the recovered covariance of the truth around the MAP estimate. 
To do this, we compare and match Eqns.~\ref{eq:var_post_overxandS} \&~\ref{eq:var_dist}, to derive a Bayesian posterior parameterised by $m_\textrm{match}$ that gives a posterior distribution that, averaged over multiple trials, has a  model parameter covariance that matches the distribution of MAP estimates that we would get from repeating the experiment. This assumes that, for these repeated trials, ${\bm x}_0$ is drawn from a Gaussian distribution around the true cosmological model. In this case, the equation for $m_\textrm{match}$ is
\begin{equation}  \label{eq:M}
  m_\textrm{match} = n_\theta+2+
    \frac{n_s-1+B(n_d-n_\theta)}{1+B(n_d-n_\theta)}\,.
\end{equation}


The resulting values of $m_\textrm{match}$ are compared in Fig.~\ref{fig:M} for a range of values of $n_s$, $n_d$, and $n_\theta$. As can be seen, $m_\textrm{match}$ tends towards the \citet{SH16} solution $m=n_s$ for large values of $n_s$. However there are differences, especially when $n_s\sim n_d$ and the posterior is more influenced by the prior than when many more simulations are available. We note that this is derived under a number of assumptions, particularly that of a linear model, and so this is still an approximation to a true matched posterior given a more complicated shape and non-linear model dependence. In particular, we caution that the moment-matching prior is not invariant to reparametrisation. We find that the exponent for $n_d=2, n_\theta=1$ is very close to that derived from the right-Haar prior (based on Cholesky decomposition of the covariance matrix), which has some exact matching properties for Gaussian variables \citep{SB06}.

\begin{figure}
  \centering
  \resizebox{0.44\textwidth}{!}{\includegraphics{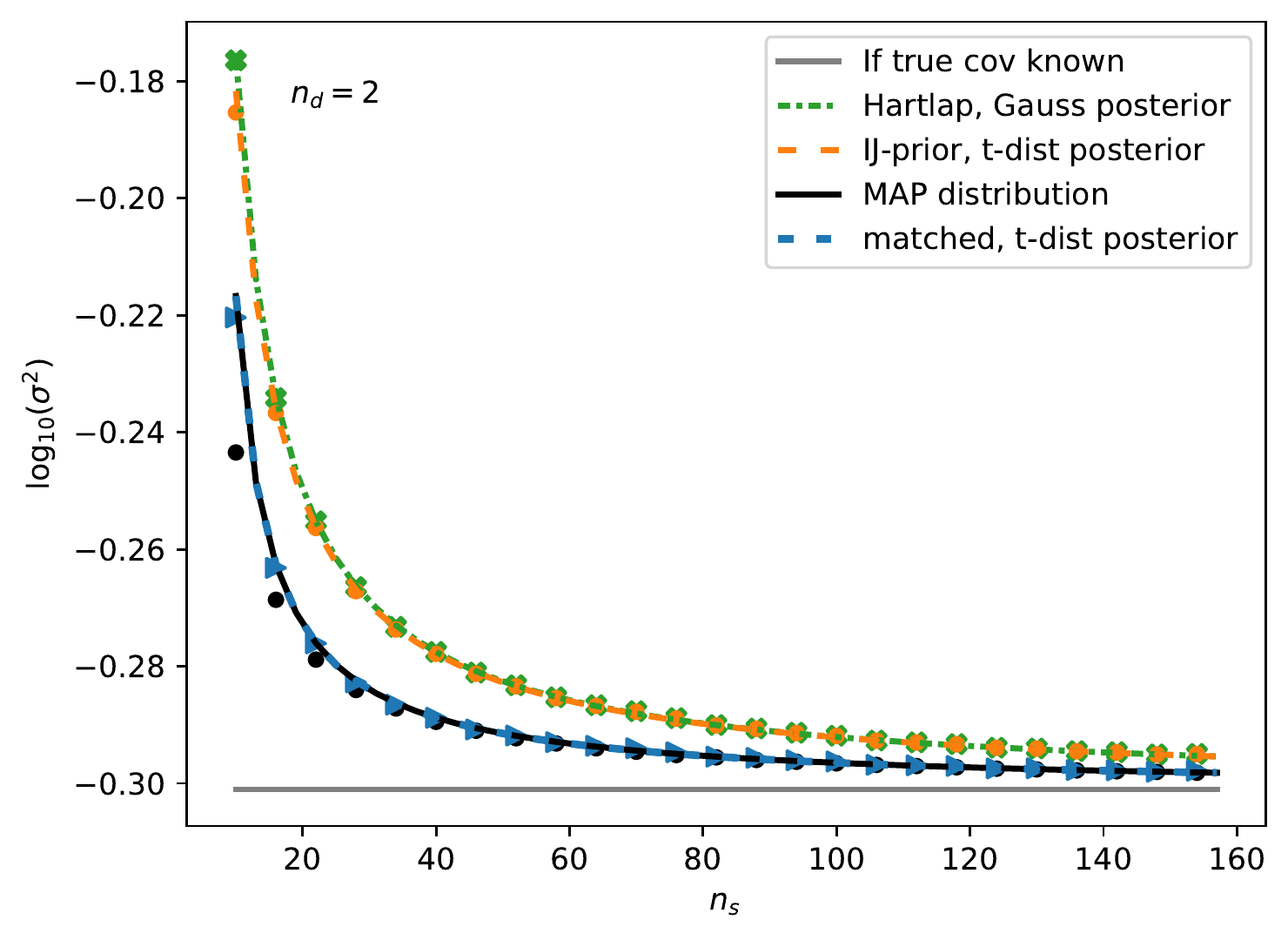}}
  \resizebox{0.44\textwidth}{!}{\includegraphics{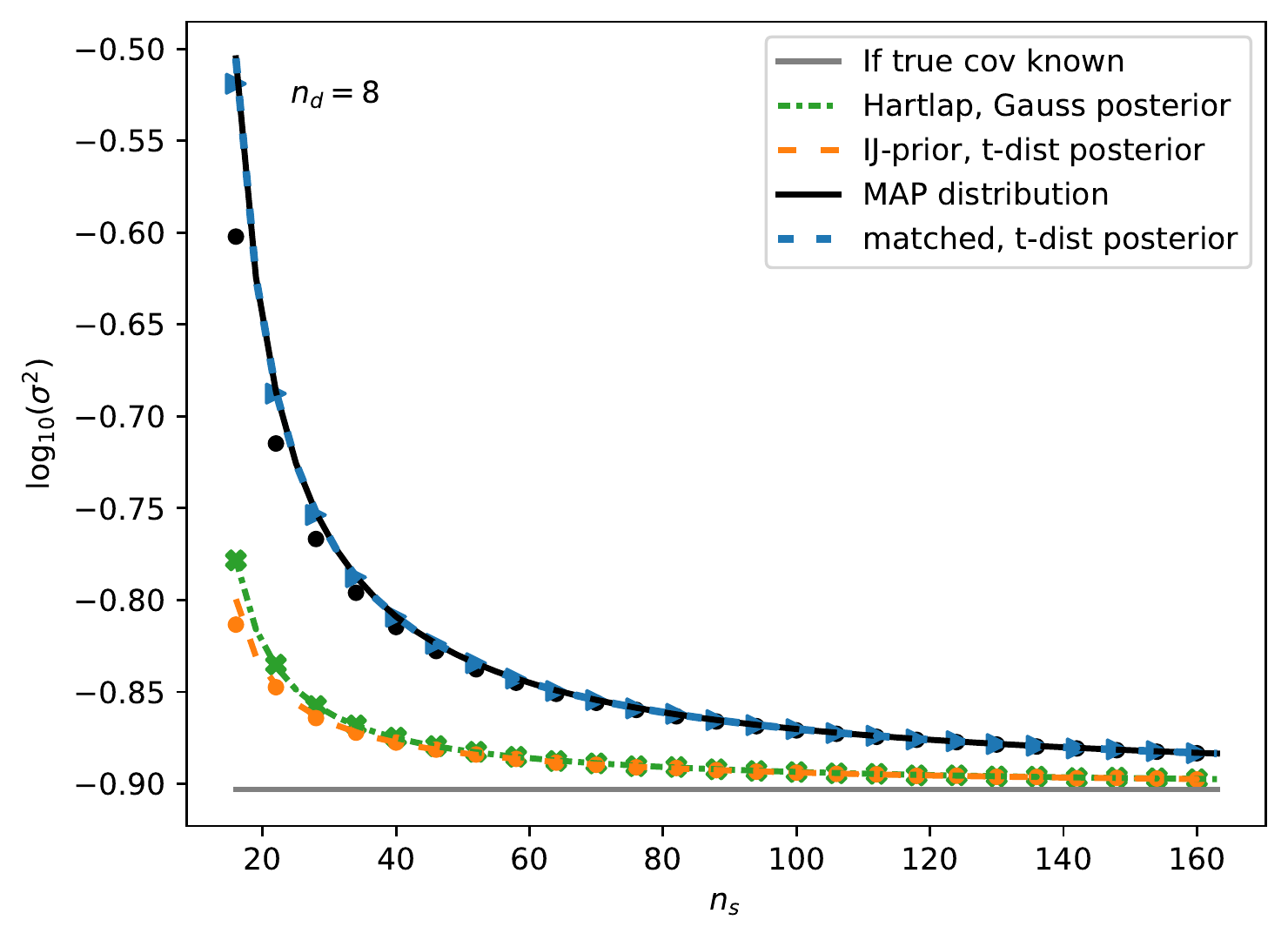}}
  \resizebox{0.44\textwidth}{!}{\includegraphics{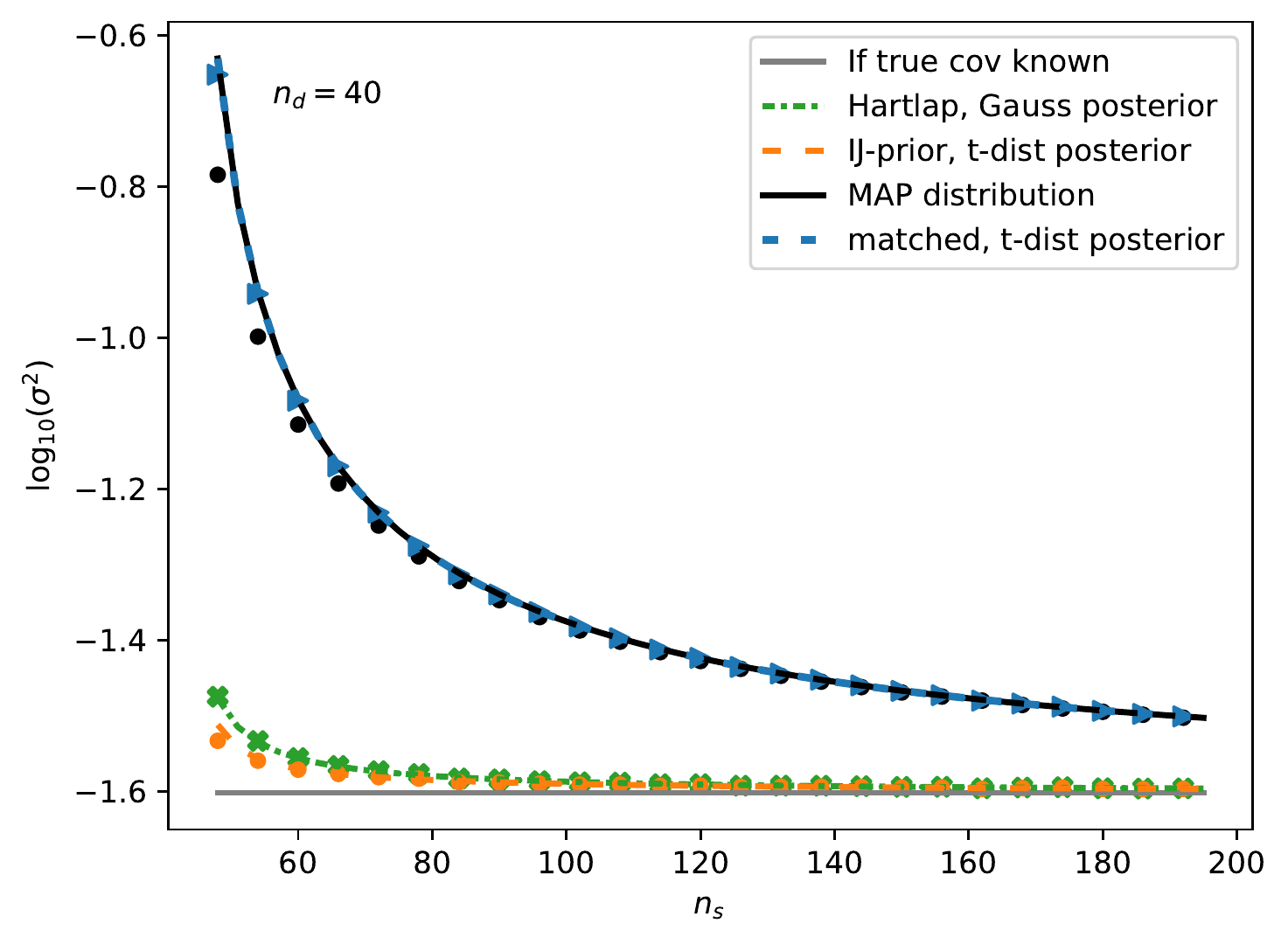}}
  \caption{Average recovered model parameter variance calculated in different ways when fitting the mean $\bar{\mu}$, to a set of correlated Gaussian data. The \corr{grey} solid line shows the result that would have been obtained from the posterior if we had known the data covariance matrix perfectly, taking the confidence interval as the root of the variance. The black solid line (model from \citealt{DS13}) and solid black points (Monte Carlo measurements) show the root of the variance determined from the distribution of recovered values. The difference is likely due to the perturbative nature of the derivation of the expectation. \corr{Green and orange} lines and \corr{symbols} show that we recover the similar distributions from both Hartlap-corrected Gaussian and t-distribution posteriors calculated with an independence Jeffreys prior as advocated in \citet{SH16}. This can easily be understood as the posteriors have the same variance. The \corr{blue dashed} lines and \corr{triangles} show the result of using a t-distribution posterior with $m_\mathrm{match}$, corresponding to Eq.~\ref{eq:M}. As can be seen, this prior is able to match the variance recovered by integrating under the posterior averaged over our realisations, with the scatter recovered from MAP estimates. \label{fig:var}}
\end{figure}

\section{Testing with a simple mean fitting model}  \label{sec:MC_test}

The resulting covariance matrix for the model parameters is tested and explored by considering a simple model - that of fitting a mean value to correlated data. We create Monte Carlo simulations that step through different realisations of the data (Gaussian distributed with covariance $\Sigma$, chosen for convenience to be the identity matrix) and analysed with covariance matrix $S$ drawn from a Wishart distribution (degrees of freedom $n_s-1$ and scale matrix $\Sigma$). Inferences are made about credible intervals assuming different choices for the posterior, and the derived estimates of the  model parameter covariances are then averaged over multiple realisations. Averaging over realisations of the data and covariance matrix $S$ in this way most naturally follows the ethos behind the derivation in Section~\ref{sec:Var}. We also record the MAP estimates for the model, and consider the distribution of these MAP estimates around the true values and measure the variance of this distribution.

We create large numbers of realisations of data ${\bm x}_0$ and covariance matrices $S$ and then fit to each assuming different expressions for the posterior. For each covariance matrix $S$, we create $n_s$ different versions of ${\bm x}_0$, and we create $100\,000$ different covariance matrices. To speed up these calculations we use analytic marginalisation over the posterior for each $S$, as outlined in Appendix~\ref{app:marginalisation}, rather than numerically integrating under the posterior for each, and use library routines to calculate realisations of Wishart matrices. We still use Monte Carlo results for different values of $S$, and the distribution of MAP parameters. Results are shown in Fig.~\ref{fig:var}, which shows that, as expected, we can choose a prior to match the  model parameter covariance recovered from the posterior to that calculated in a frequentist style approach where we look at the spread of recovered MAP estimates. Given that this derivation best matched the setup of the Monte Carlo simulations, using $m_\textrm{match}$ provides an excellent fit to the numerical results. 

\section{Testing against a non-linear model}
\label{sec:DES_like}

To test the performance of the different posterior distributions discussed in Sections~\ref{sec:bayesian} and \ref{sec:matching} in a realistic cosmological setting we adopt a mock experiment as also considered by \citet{Friedrich18}. They simulated a tomographic cosmic shear data vector including auto- and cross-correlations of $\xi_\pm$ in 5 source redshift bins on a survey area of 5000 $\deg^2$ (hence mimicking 5-year data of the Dark Energy Survey, cf.\ their table 1 for details). Overall this data vector contains $450$ data points. Around a true data vector computed at a cosmology with $(\Omega_m , \sigma_8, w_0) = (0.3156, 0.831, -1)$ we draw $1000$ Gaussian random realisations assuming a theoretical covariance matrix derived using the halo model to describe non-linear clustering. Here $\Omega_m$ is the present day cosmological matter density, $\sigma_8$ is the rms density fluctuations in spheres of radius 8\,h$^{-1}$Mpc, and $w_0$ is the Dark Energy equation of state parameter. Both the covariance calculation and subsequent analyses of the mock data vectors are carried out with the CosmoLike toolkit \citep{Krause:2016jvl}.

\begin{figure}
  \centering
  \resizebox{0.45\textwidth}{!}{\includegraphics{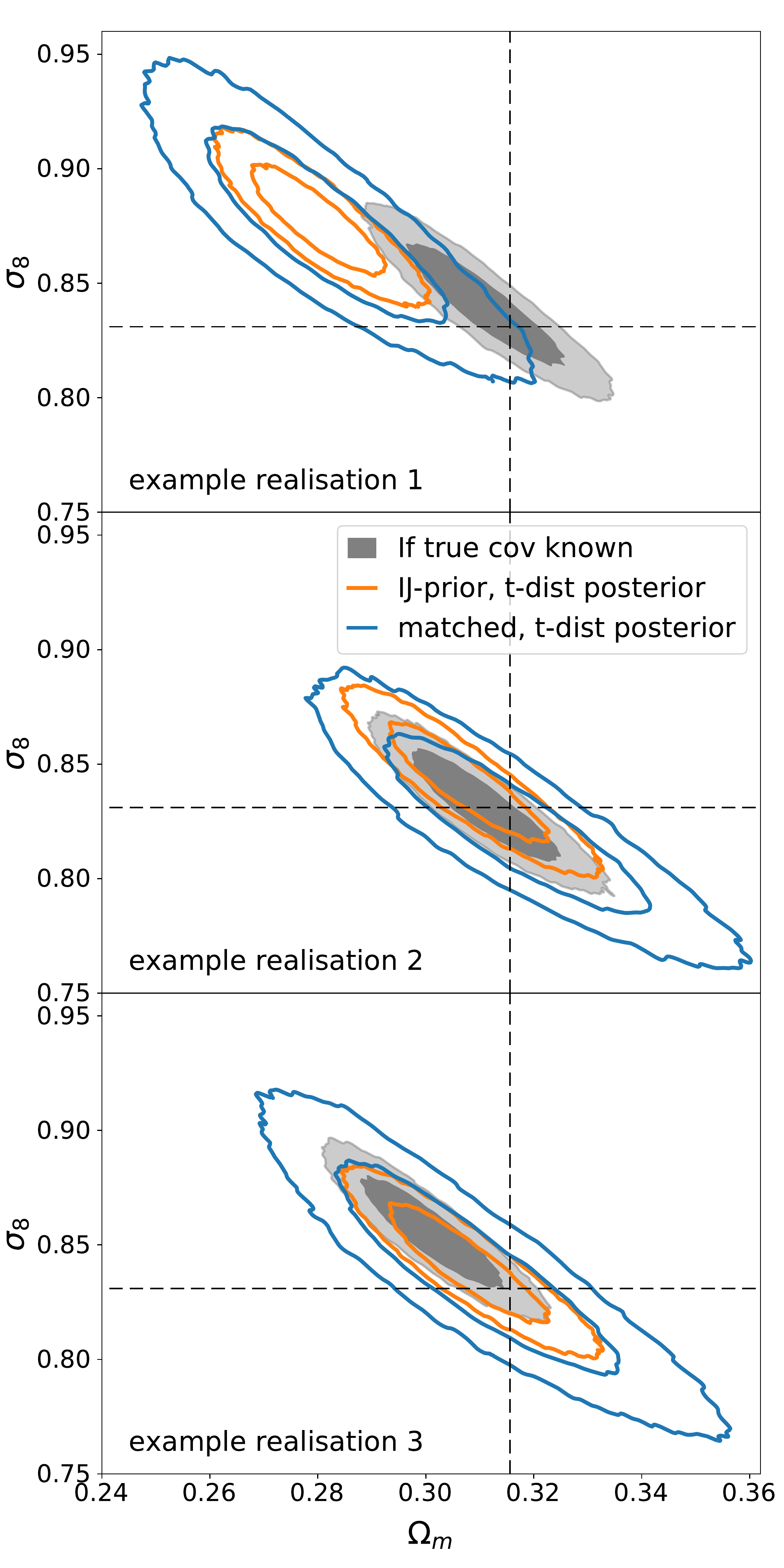}}
  \caption{\corr{Contours containing 68\% and 95\% probability, marginalising under the} posterior in the $\Omega_m$-$\sigma_8$ plane, obtained from realizations of DES-like weak lensing data vectors. Each panel is for a different random set of data ${\bm x}_0$ and covariance $S$ drawn from Gaussian and Wishart distributions respectively. The relevant parameters of this run for the posterior are $n_s=650$, $n_d=450$, and $n_\theta=4$. Contours are shown calculated using the true covariance matrix with a Gaussian posterior (\corr{grey shading}), and two versions of the t-distribution posterior, one with $m=n_s$ as derived using an independence Jeffreys prior for the true covariance as in \citet{SH16} (\corr{orange}), and one using a covariance-matching prior derived for linear models (\corr{blue}). \corr{The dashed lines mark the expected values of both parameters.} \label{fig:des-like}}
\end{figure}

In Fig.~\ref{fig:des-like} we show marginalised posterior constraints in the $\Omega_m$-$\sigma_8$ plane obtained from the first three of our random realisations using different posteriors. The \corr{grey shaded} contours were obtained using the true analytic covariance that was also used to draw our mock data vectors. The orange contours assume that there is a covariance estimate from $650$ simulations (i.e.\ 200 more \corr{than} data points) and that this estimate is used in the posterior of \citet{SH16} to obtain the constraints (we draw a new covariance estimate for each data vector from a Wishart distribution). Note that \corr{all} contours within each individual panel of Fig.~\ref{fig:des-like} are derived from the same data vector realisations. Despite that, there is a noticeable additional scatter between the two sets of contours - this is exactly the effect of additional scatter of MAP estimates due to noisy covariance estimates described by \citet{DS13}. The blue contours are the modified version of the posterior with a $m_\textrm{match}$ prior chosen to to match this additional scatter.

\begin{figure*}
  \centering
  \resizebox{0.9\textwidth}{!}{\includegraphics{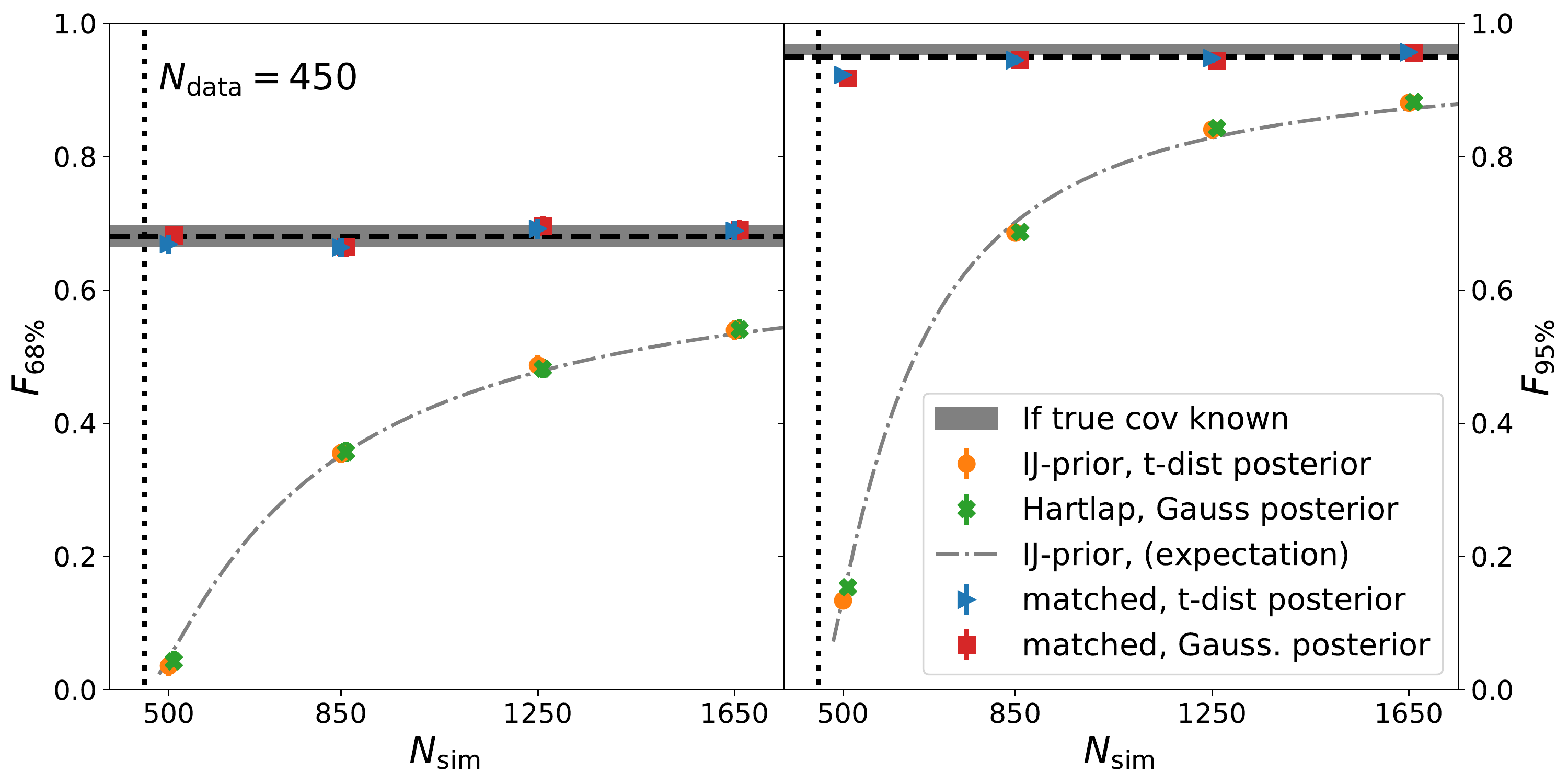}}
  \caption{Comparing how often the true cosmology underlying our numerical experiment of Sec.~\ref{sec:DES_like} is found inside the 68\% (left panel) and 95\% (right panel) credible regions when using different covariance matrices and different posterior distributions. Note that we are comparing how well a credible region works as a confidence interval, and hence this test is not fair. The grey band in each panel assumes that the true covariance is known, in which case, with the problem being considered the credible interval also works as a confidence interval. \corr{The width of the band indicates the expected credible interval containing a coverage probability of 68\% from 1000 realisations, assuming a binomial distribution for the number of successes. Hence the horizontal black dashed line, which indicated the expected value does not have to lie in the middle of this band.} The \corr{green} crosses represent the \corr{common} approach of a Gaussian likelihood with Hartlap corrected precision matrix \corr{for different numbers of simulations used to estimate the covariance matrix (x-axis in both panels). The vertical dotted line marks $n_s=n_d$.} The \corr{orange} dots use the independence Jeffreys prior advocated by \citet{SH16} and the resulting t-distribution instead of the Hartlap corrected Gaussian likelihood. The \corr{blue} triangles show the coverage achieved with a matched prior that uses Eq.~\ref{eq:M} to compute the exponent $m$, and the red squares show the coverage obtained from simply rescaling the Gaussian log-likelihood in the manner advocated by \citet{Percival14}, as in Eq.~\ref{eq:Sprime}. The dash-dotted line show the coverage that is expected for the t-distribution posterior calculated using the independence Jeffreys prior. 
  }
  \label{fig:des-like_coverage}
\end{figure*}

To assess the performance of our matched prior more quantitatively we run our Markov Chain Monte Carlo routine to explore the posteriors around all 1000 random \corr{realisations} of our data vector. Fig.~\ref{fig:des-like_coverage} compares how often the true cosmology underlying our numerical experiment is located inside the 68\% (left panel) and 95\% (right panel) confidence regions of the full 3-dimensional parameter space when using different covariance matrices and different posterior distributions. Here we are considering the credible intervals derived from our Bayesian analysis work as frequentist confidence intervals. The grey band in each panel assumes that the true covariance is known. The \corr{green} crosses represent the \corr{commonly used} approach of a Gaussian likelihood with Hartlap corrected precision matrix as estimated from different numbers of simulations (x-axis in both panels). The \corr{orange} dots use the independence Jeffreys prior advocated by \citet{SH16} and the resulting t-distribution instead of the Hartlap-corrected Gaussian likelihood. The \corr{blue triangles} show the coverage achieved with a matched prior that uses Eq.~\ref{eq:M} to compute the exponent $m$. This likelihood indeed manages to achieve coverage factions of approximately 68\% and 95\% respectively. The red \corr{squares} show the coverage obtained from simply re-scaling the Gaussian log-likelihood in the manner advocated by \citet{Percival14}, which is also close to 68\% and 95\% respectively. The dash-dotted line show the coverage that is expected for the standard Gaussian likelihood based on the calculations of \citet{DS13}.


\section{Summary} \label{sec:summary}

Our suggested way forwards is quite simple - in situations where the covariance matrix $S$ for Gaussian data is itself a random variable drawn from a Wishart distribution with $n_s-1$ degrees of \corr{freedom}, for example when it is constructed from $n_s$ mock samples, then we propose a frequentist matching prior that is uniform in ${\bm \mu}$ and depends on $\Sigma$ as $|\Sigma|^{-(m-n_s+n_d+1)/2}$, leading to a posterior 
\begin{equation}
    f({\bm \mu}|{\bm x}_0,S) \propto
    \left[1+\frac{\chi^2}{(n_s-1)}\right]
        ^{-\frac{m}{2}}\,,
\end{equation}
where
\begin{equation}
    \chi^2 = ({\bm x}_0-{\bm \mu})\corr{^T}S^{-1}({\bm x}_0-{\bm \mu})\,.
\end{equation}
The power law index $m$ is given by Eq.~\ref{eq:M}, and repeated here for completeness
\begin{eqnarray}
  m&=&n_\theta+2+
    \frac{n_s-1+B(n_d-n_\theta)}{1+B(n_d-n_\theta)}\,,\\
      B&=&\frac{(n_s-n_d-2)}{(n_s-n_d-1)(n_s-n_d-4)}\,,
\end{eqnarray}
where $n_d$ is the number of data points and $n_\theta$ the number of parameters. This will lead to credible intervals that can also be interpreted as confidence intervals with approximately the same coverage probability. Note that this expression does not require any extra factors of $h$, or other terms - i.e. $S$ is the approximate covariance matrix, and $S^{-1}$ its inverse. This enables a Bayesian analysis, with a matching prior designed with this frequency-matching property. In general, this procedure increases the model parameter credible intervals compared with those derived from the more usual independence Jeffreys prior on the true data covariance, and therefore can be considered a more conservative choice for making deductions from data.

If the reader prefers to approximate the posterior using a Gaussian distribution, then rather than inverting $S$ or $hS$, the matrix $(S')^{-1}$ to be used when calculating $\chi^2$ should be the inverse of
\begin{equation}  \label{eq:Sprime}
    S' = \frac{(n_s-1)[1+B(n_d-n_\theta)]}{n_s-n_d+n_\theta-1}S\,,
\end{equation}
which matches the method proposed in \citet{Percival14}, replacing one of the approximations used there with an exact expression. To derive this, consider the factor by which we must multiply Eq.~\ref{eq:invFish-Gauss} to obtain Eq.~\ref{eq:var_dist} - matching the model parameter covariance expected from integrating under the posterior with that from the distribution of MAP solutions.

Both the Gaussian approximation and our preferred t-distribution solution give  model parameter covariances that are very similar to the suggestion of \cite{Friedrich18} when $n_\theta$ is small. They proposed multiplying the \citet{SH16} posterior by the \citet{DS13} factor of $1+B(n_d-n_\theta)$. To see the empirical similarity, note that the \citet{SH16} posterior gives a covariance for the distribution of $\bm\mu$ around ${\bm x}_0$ of $hS$, and compare Eq.~\ref{eq:Sprime} to $hS\times[1+B(n_d-n_\theta)]$.

\section{Conclusions}  \label{sec:discussion}

The primary result in our paper is presented in Section~\ref{sec:summary}, which provides a frequentist-matching prior: i.e. the exponent in a power law prior on the determinant of the true data covariance matrix required to yield a posterior model parameter covariance matching the distribution of true parameter values with respect to maximum likelihood estimates (and vice-versa for the linear models we consider). Our analysis lies at the interface between Bayesian and frequentist analyses: allowing an analysis that results in multiple interpretations of the same parameter intervals with the same probability. In order to derive this, we have assumed a linearised model, but have demonstrated broader applicability using a realistic non-linear model fit. Note that, in general, our results will not be valid for arbitrary non-linear models or reparametrisations. The use of this formalism for parameter inference when the covariance matrix is itself approximate offers a way to satisfy scientists whose intuition is based on frequentist style measures and those who wish for the analysis to be Bayesian in construct (which is often simpler for practical application). 

We initially considered an independence Jeffreys prior on the true covariance matrix, as advocated in \citet{SH16}. We showed that this leads to a posterior with covariance around the model parameters that matches that assuming a Gaussian posterior after scaling the data covariance matrix by the Hartlap factor. The derived model parameter covariance does not match that from the distribution of MAP estimates found by \citet{DS13}, which is understandable given that they are calculating different distributions. We have considered alternative priors that are powers of the determinant of the true covariance matrix and which yield posteriors with frequentist coverage, at least at the level of covariance of the distributions. Using this allows the interpretation of credible intervals as confidence intervals with approximately the same probability. Because of the choice of a power-law prior, the posteriors of interest have the form of a multivariate t-distribution. For this form, the distribution of the posterior around the MAP estimate depends on the specific data realisation - this can clearly be seen in Eq.~\ref{eq:var_post_givenxandS}. In comparison, for a Gaussian posterior, the distribution around the MAP estimates is independent of the data and depends only on the data covariance matrix $S$. This complicates the matching. We therefore consider the recovered  model parameter covariance averaged over a set of data: here the distribution of that data matters. Formally, we calculate the frequentist coverage probability for a set of credible intervals, with a view to matching this probability to that from the distribution of MAP solutions.

Although we have a t-distribution posterior, the distribution of data is Gaussian, and so we cannot directly use either the t-distribution Fisher matrix (this led to expected covariance on model parameters as in Eq.~\ref{eq:var_fish}), or integrate under the posterior assuming the data is distributed according to a multivariate t-distribution (leading to Eq.~\ref{eq:cov-theta-tdist-m}). Instead, we have to consider the Gaussian distribution of data when determining the average  model parameter covariance that would be recovered from the posterior after repeated trials (giving Eq.~\ref{eq:var_post_overxandS}). We also note that this dependence on the data complicates data compression: the credible intervals recovered from compressed data do not necessarily match those recovered from the full data even for linear models where the compression is optimally performed to give the same MAP estimates (see Appendix~\ref{sec:Compress}).


The prior that we advocate depends on the properties of the data and the problem, particularly $n_s$, $n_d$ and $n_\theta$. Having priors that depend on the expected form of the posterior is quite common (although they should obviously not depend on the actual data observed), especially in the objective Bayesian approach (see \citealt{HS18} for an application to cosmology), so we do not see this as a fundamental problem, although it does conflict with the Bayesian notion of the prior as an expression of the state of knowledge before the experiment is performed.

One might also worry that our matching criterion is, in a sense, linking the posterior and properties of the data that depend on the likelihood. But the posterior should answer the question of what is the truth given the data, while the likelihood considers the data given the truth. These are fundamentally different things, and so why are we matching posterior and likelihood widths? If we compare the covariance inherent in the likelihood and the posterior for multivariate Gaussian distributions, then we might consider an approximate link where $\sigma^2_{\rm post}=\sigma^2_{\rm like}+\sigma^2_{\rm prior}$. This would be exact if all distributions were Gaussian, or we were working in the Gaussian limit. In this limit, the standard prior on the covariance used in the posterior directly adds to the covariance we assume for our experimental result. Translating through to model parameters, both contributions still contribute. So we see that the prior choice is related to the credible interval quoted for experimental measurements and forms the link between posterior and likelihood. A prior is chosen such that it does not change this covariance, and so in this sense our matching prior is an uninformative prior for the model parameters. 

Using the multivariate t-distribution posterior makes the analysis attractive in a Bayesian sense, as it matches the problem with fewer approximations. In general, approximating the posterior as Gaussian has a relatively small effect on the posterior surface for $1\sigma$ and $2\sigma$ intervals, and in the examples we have considered less so than the choice of prior (see Appendix~\ref{sec:intervals}). Even so, we recommend using the multivariate t-distribution with the revised prior as this represents a consistent Bayesian approach. Moreover, the tail probabilities can be much greater than those of the equivalent gaussian, which can be in error when tensions between datasets are considered. In this case, we need to be careful about the interpretation of $N\sigma$ confidence intervals, as discussed in Appendix~\ref{sec:intervals}. For those that cannot contemplate a posterior with a form other than Gaussian, we have included the alternative correction to use instead of the Hartlap factor for an approximate Gaussian posterior in Section~\ref{sec:summary}.

When $n_\theta=n_d$, Eq.~\ref{eq:M} gives that $m=n_s+n_d+1$, and the prior reduces to $|\Sigma|^{-(n_d+1)}$. For this prior, the covariance of the posterior distribution as given in Eq.~\ref{eq:var_tdist_m} reduces to $S$. From the properties of the Wishart distribution, this has expected value $\Sigma$ matching the covariance of the frequentist distribution from which the data were assumed to be drawn. Note that no factor of $h$ is required in the posterior, or in the Gaussian approximation to get this result. To understand why not, note that the rationale often used to justify using a Gaussian posterior based on a covariance $hS$ (i.e. including a factor $h$) is that the inverse matrix $S^{-1}$ is a biased estimate of $\Sigma^{-1}$, and this is corrected by using $hS$ rather than $S$. Thus the argument goes that we should use $hS$ in the posterior. However, we should consider that the model parameter covariance derived from the posterior is biased in the opposite way requiring an extra factor $h^{-1}$ following the same rationale. To see this, consider Eq.~\ref{eq:MSinvM}, which shows that the model parameter covariance from a set of repeated trials each with a different $S$ (with no $h$ factor) is Wishart distributed with expectation given by a function of $\Sigma$. Where $n_\theta=n_d$, and we fit for the values of $\bm\mu$, the derived covariance reduces to $S$ with expectation $\Sigma$, matching that we would expect given the Gaussian distribution of the data. Including $h$ would have biased our errors compared to this expected value. Thus, explicitly including the Hartlap factor in a posterior to correct for a bias in $S^{-1}$ is not just wrong from a Bayesian standpoint, but the standard rationale for its application misses a crucial step. Our proposed posterior consistently corrects for any potential biases due to having skewed distributions without any need for extra ad-hoc factors.

Finally we note that we form a matching prior based on the recovered model parameter covariance and not the distribution, as is more standard in statistical analyses. We do this because the covariance of the posterior distribution for model parameters offers a simple way to match the "width" of two distributions, and that we can determine simple results for a power-law prior where we only have one degree of freedom and so only one degree of matching is possible. An extension to this work would be to consider varying the form of the prior beyond a simple power-law of the determinant of the true data covariance matrix to better match the shape of the posterior, in line with the more standard matching criterion used in statistics. We could also have directly compared credible intervals and confidence intervals - i.e. averaged over $\sigma$ rather than the model parameter covariance where necessary, but we do not expect that this would change our results significantly compared with our chosen matching criterion based on covariance. 

\section*{Acknowledgements}

WJP acknowledges useful conversations with Michael Matesic and En Long regarding speeding up running the Monte Carlo realisations. We thank Daniel Farewell for helpful comments on an early version of the draft and Tim Eifler for useful comments on the draft and for providing the CosmoLike toolkit. \corr{We thank the referee, James Buchanan, for their careful review of the paper and for the corrections and suggestions provided.}

Research at Perimeter Institute is supported in part by the Government of Canada through the Department of Innovation, Science and Economic Development Canada and by the Province of Ontario through the Ministry of Colleges and Universities.

This research was enabled in part by support provided by Compute Ontario (www.computeontario.ca) and Compute Canada (www.computecanada.ca).

\section*{Data Availability}

No data was used in this paper, which is theoretical in nature. 



\bibliographystyle{mnras}
\bibliography{matching-prior-refs} 




\appendix

\section{Multivariate distributions}  \label{sec:mult-dists}

Some multivariate distributions with data dimension $n_d$ are listed here for reference:

\textbf{The Wishart distribution}
\begin{equation}  \label{eq:wishart}
  f_W(S|R,\nu)=
  \frac{|S|^\corr{\frac{\nu-n_d-1}{2}}
    \exp\left[-\frac{1}{2}Tr(R^{-1}S)\right]}
  {2^{\frac{\nu n_d}{2}}|R|^{\frac{\nu}{2}}\Gamma_\corr{n_d}\left(\frac{\nu}{2}\right)}\,,
\end{equation}
where $\nu$ is the degrees of freedom, and $R$ the scale matrix. \corr{The mean is $E[S]=\nu R$, and the variance is Var$[S_{ij}]=\nu[R_{ij}^2-R_{ii}R_{jj}]$.}

\textbf{The inverse Wishart distribution}
\begin{equation}  \label{eq:iwishart}
  f_{W^{-1}}(R|S,\nu)=\frac{|S|^{\frac{\nu}{2}}|R|^{-\frac{\nu+n_d+1}{2}}
    \exp\left[-\frac{1}{2}Tr(R^{-1}S)\right]}
  {2^{\frac{\nu n_d}{2}}\Gamma_{n_d}\left(\frac{\nu}{2}\right)}\,,
\end{equation}
where $\nu$ is the degrees of freedom, and $S$ the scale matrix. \corr{The mean is $E[R]=S/(\nu-n_d-1)$.}

\textbf{The multivariate Normal or Gaussian distribution} written in a form
using the Trace operator
\begin{equation}  
  f_N({\bm x}_0|{\bm \mu},R) = 
  \corr{(2\pi)^{-\frac{n_d}{2}}|R|^{-\frac{1}{2}}}\exp\left[-\frac{1}{2}Tr
    \left(R^{-1}({\bm x}_0-{\bm \mu})({\bm x}_0-{\bm \mu})^T\right)\right]\,,
\end{equation}
with mean $E[{\bm x}_0]={\bm \mu}$ and variance Var$[{\bm x}_0]=R$.

\textbf{The multivariate t-distribution}
\begin{displaymath}
  f_{t,\nu}({\bm x}_0|{\bm \mu},R) =
  \frac{\Gamma[(\nu+n_d)/2]}{\Gamma(\nu/2)(\nu\pi)^{n_d/2}|R|^{1/2}}
\end{displaymath}
\begin{equation}  \label{eq:tdist}
    \hspace{1cm}\times
    \left[1+({\bm x}_0-{\bm \mu})\corr{^T}(\nu R)^{-1}({\bm x}_0-{\bm \mu})\right]^{-\frac{\nu+n_d}{2}}\,,
\end{equation}
where $\nu$ is the degrees of freedom, and $R$ the scale matrix. \corr{The mean is $E[{\bm x}_0]={\bm \mu}$, and the variance is Var$[{\bm x}_0]=\frac{\nu}{\nu-2}R$.}

\section{Perturbative based approach for expressions involving the covariance of Wishart-distributed matrices}  \label{sec:pert}

In this Appendix, we consider the perturbation based approach to understanding the biases involved in a statistical analysis of data when the covariance matrix itself is a random variable $S$. To do this, we use the expressions between estimated and true covariance matrix as provided by \citet{Taylor13}. Let $(hS)^{-1} = \Sigma^{-1}+\Delta_{\Sigma^{-1}}$. As $S$ is drawn from a Wishart distribution, the errors $\Delta_{\Sigma^{-1}}$ can be written
\begin{equation} \label{eq:cov-wish}
  \langle (\Delta_{\Sigma^{-1}})_{ab}(\Delta_{\Sigma^{-1}})_{cd}\rangle_S
  = A\Sigma^{-1}_{ab}\Sigma^{-1}_{cd}
  +B(\Sigma^{-1}_{ac}\Sigma^{-1}_{bd}+\Sigma^{-1}_{ad}\Sigma^{-1}_{bc})\,,
\end{equation}
where
\begin{eqnarray}  \label{eq:AB}
  A&=&\frac{2}{(n_s-n_d-1)(n_s-n_d-4)}\,,\nonumber\\ 
  B&=&\frac{(n_s-n_d-2)}{(n_s-n_d-1)(n_s-n_d-4)}\,. 
\end{eqnarray}

First, we consider a perturbative expansion of $F_S^{-1}=\corr{h^{-1}}(F_\Sigma+\Delta_F)^{-1}$, with $\Delta_F$ defined as a standard Gaussian Fisher matrix with inverse covariance $\Delta_{\Sigma^{-1}}$ as required in Section~\ref{sec:Fish}. Expanding this, and taking the expected value, the first order terms in $\Delta_F$ tend to zero (as $(hS)^{-1}$ is an unbiased estimator of $\Sigma^{-1}$), and so we are only interested in the second order term in $\Delta_F$, which can be written
\begin{equation} \label{eq:Ftinv2}
 \left.\langle(F_\Sigma+\Delta_F)^{-1}\rangle_S\right|_{s.o.} = F_\Sigma^{-1}\Delta_F\,F_\Sigma^{-1}\Delta_F\,F_\Sigma^{-1}\,.
\end{equation}
Putting the relationships given in \corr{Eq.~\ref{eq:cov-wish}} into Eq.~\ref{eq:Ftinv2}, we find that
\begin{equation}  \label{eq:Finv-pert}
  \langle F_S^{-1}\rangle_S \simeq   h^{-1}\left[1+A+B(n_\theta+1)\right]
    F_\Sigma^{-1}\,.
\end{equation}
The calculation of the inverse Fisher matrix averaged over $S$ using this perturbation based approach was performed in \citet{Percival14} for the Gaussian Fisher matrix. As shown in the derivation leading to Eq.~\ref{eq:var_fish}, this expression does not have to be solved perturbatively as an exact solution is possible. The non-perturbative solution is given in Eq.~\ref{eq:invFish-Gauss}.

The next expression that we wish to understand perturbatively is $\langle F_S^{-1}Tr[S^{-1}\Sigma]\rangle_S$, as required in \corr{Section~\ref{sec:var_post_gauss}} and given in Eq.~\ref{eq:pert-mid}. The \corr{expression for which we are taking the expectation can be written}
\begin{equation}
\corr{(F_S^{-1}Tr[S^{-1}\Sigma])}_{\alpha\beta}=
 [F_S^{-1}]_{\alpha\beta} S^{-1}_{ab}
    \Sigma_{ab}\,.
\end{equation}
The second order term from $F_S^{-1}$ is given by Eq.~\ref{eq:Finv-pert}, leading a term $n_d[1+A+B(n_\theta+1)]$, with the factor $\corr{n_dh}$ coming from the summation over the term $S^{-1}_{ab}\Sigma_{ab}$. There is also a second order cross term from $F_S^{-1}$ and $S^{-1}_{ab}$, which gives $-[n_dA+2B]$. Adding these together, we find the result in Eq.~\ref{eq:pert-mid}.

To approximate the expression in Eq.~\ref{eq:pert-end}, note that there are eight possible ways that we can have pairs of $\Delta_{\Sigma^{-1}}$ in 
\begin{equation}  \label{eq:dist1}
    [F_S^{-1}]_{\alpha\beta}
    \frac{d\mu_a}{d\theta_{\alpha'}}
    S^{-1}_{ab}
    \Sigma_{bc}
    S^{-1}_{cd}
    \frac{d\mu_d}{d\theta_{\beta'}}
    [F_S^{-1}]_\corr{\beta'\alpha'}\,,
\end{equation}
with one at second order from each $F_S^{-1}$, the cross pair between the two $F_S^{-1}$ and the cross pair from the two $S^{-1}$, and four cross pairs between $F_S^{-1}$ and $S^{-1}$. Treating each in turn and expanding using \corr{Eq.~\ref{eq:cov-wish}} leads to the result in Eq.~\ref{eq:pert-end}.

Finally, we note that the expression in Eq.~\ref{eq:var_dist} can be derived similarly. To see this, note that there are eight possible ways that we can have pairs of $\Delta_{\Sigma^{-1}}$ in
\begin{equation}  \label{eq:dist2}
    \langle \hat{\bm\theta}_\alpha\hat{\bm\theta}_\beta\rangle_{x,S}
    = [F_S^{-1}]_{\alpha\alpha'}
    \frac{d\mu_a}{d\theta_{\alpha'}}
    S^{-1}_{ab}
    \Sigma_{bc}
    S^{-1}_{cd}
    \frac{d\mu_d}{d\theta_{\beta'}}
    [F_S^{-1}]_\corr{\beta'\beta}\,,
\end{equation}
similar to the expansion of Eq.~\ref{eq:dist1}. These expressions are different - for example in the limit as $hS\to\Sigma$, Eq.~\ref{eq:dist1} tends towards $n_\theta(F_\Sigma^{-1})_{\alpha\beta}$, while Eq.~\ref{eq:dist2} tends towards $(F_\Sigma^{-1})_{\alpha\beta}$. Treating each of the eight possible combinations of two $\Delta_{\Sigma^{-1}}$ separately, expanding using \corr{Eq.~\ref{eq:cov-wish}} and summing the terms gives the result in Eq.~\ref{eq:var_dist}, which was the primary result of \citet{DS13}.

\section{Compressing the data}  \label{sec:Compress}

The effect of a linear compression of the data on model parameter inference can be considered using a property of the multivariate t-distribution. For some $n_c\times n_d$ matrix $M$, assuming that 
\begin{equation}
  f({\bm\mu}|{\bm x}_0,S) = 
  f_{t,m-n_d}\left(\corr{\bm\mu}\left|{\bm x}_0,\frac{n_s-1}{m-n_d}S\right.\right)\,,
\end{equation}
then a property of the multivariate t-distribution is that 
\begin{equation}
    f(M{\bm \mu}|M{\bm x}_0,S) = 
    f_{t,m-n_d}\left(\corr{M{\bm \mu}}\left|M{\bm x}_0,\frac{n_s-1}{m-n_d}MSM^T\right.\right)\,.
\end{equation}
Now consider an analysis of the compressed data, where we apply a compression with $n_c=n_\theta$ and
\begin{equation}  \label{eq:compression}
    M = F_S^{-1}\frac{d{\bm\mu}}{d{\bm\theta}}^TS^{-1}\,,
\end{equation}
such that the MAP estimate $\hat{\bm\theta}=M{\bm x}_0$, and 
\begin{equation}
    MSM^T =
    F_S^{-1}\frac{d{\bm\mu}}{d{\bm\theta}}^TS^{-1}SS^{-1}
    \frac{d{\bm\mu}}{d{\bm\theta}}F_S^{-1} = F_S^{-1}\,.
\end{equation}
For data analysed with a Gaussian posterior and linear model, such a compression is sufficient in that the analysis of the reduced data gives the same inferences as those from the full data set, including the covariance on ${\bm\mu}$. Assuming a t-distribution posterior for ${\bm x}_0$, we find that the posterior for the reduced data is
\begin{equation}
    f(M{\bm\mu}|M{\bm x}_0,S) = f_{t,m-n_d}\left(\corr{M{\bm\mu}}\left|M{\bm x}_0, \frac{n_s-1}{m-n_d}MSM^T\right.\right)\,.
\end{equation}
Now, defining ${\bm\theta}'=M{\bm\mu}$, as an estimator for the MAP values, we see that
\begin{equation}
    f({\bm\theta}'|\hat{\bm\theta},S) = f_{t,m-n_d}
    \left(\corr{\bm\theta}'\left|\hat{\bm\theta},\frac{n_s-1}{m-n_d}F_S^{-1}\right.\right)\,.
\end{equation}
This gives that the covariance for ${\bm\theta}'$ is
\begin{equation} \label{eq:var_post_tdist_fg}
  \langle({\bm\theta}'-\hat{\bm\theta})({\bm\theta}'-\hat{\bm\theta})^T\rangle =
    \frac{n_s-1}{m-n_d-2}F_S^{-1}\,.
\end{equation}  
This is the covariance recovered from the compressed data as given by Eq.~\ref{eq:compression}, for a measurement of the MAP estimates $\hat{\bm\theta}$. This does not match the expression in Eq.~\ref{eq:var_post_givenxandS}, but does match the solution of Eq.~\ref{eq:cov-theta-tdist-mS} where we integrate under the posterior and then average over ${\bm x}_0$, assuming that this was drawn from a multivariate t-distribution. 

Our interpretation of this is that the linear compression of the data analysed with a t-distribution posterior does not include information about the distribution of the data around the MAP estimate, as is used in Eq.~\ref{eq:var_post_givenxandS} to determine a specific model parameter covariance for that realisation of the data. Without this extra information, compressing the data means that the model parameter covariance recovered corresponds to the average for a distribution of ${\bm x}_0$, rather than that for a particular ${\bm x}_0$ recovered if using more data. Furthermore the model parameter covariance corresponds to that recovered on average for data distributed according to a multivariate t-distribution. We therefore conclude that data compression works differently than when analysing using a Gaussian posterior for which linear compression is sufficient in terms of giving the same MAP estimate and covariance. For multivariate t-distribution posteriors, this is not the case, and additional information is used on the distribution of the data around the MAP estimate in order to determine the model parameter covariance as shown in Eq.~\ref{eq:var_post_givenxandS}. This will be considered further in future work.

\section{Interpretation of credible intervals based on $\sigma$}  \label{sec:intervals}

\begin{figure}
  \centering
  \resizebox{0.44\textwidth}{!}{\includegraphics{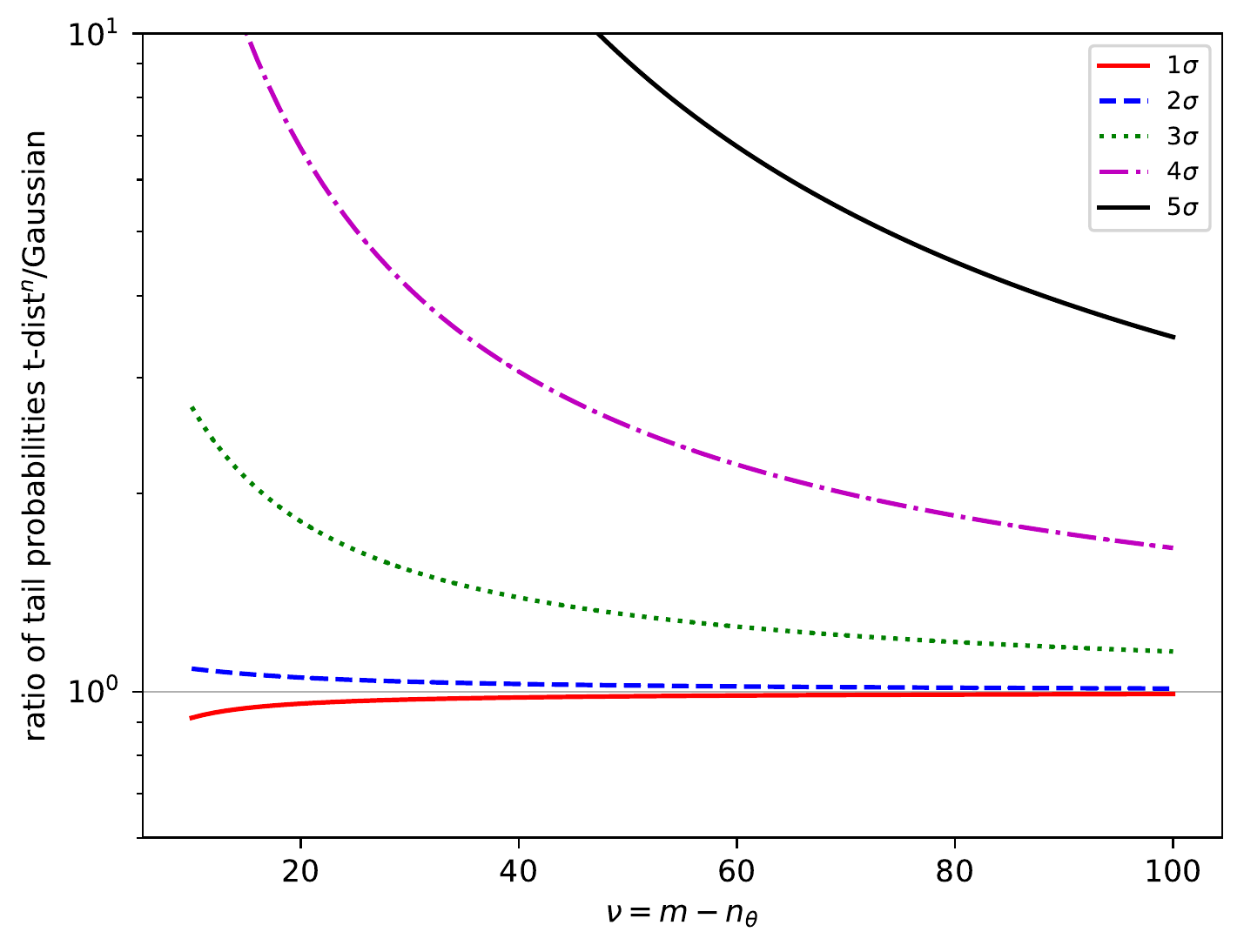}}
  \caption{The ratio of tail probabilities for t-distribution and Gaussian posteriors outside of $\pm N\sigma$ credible intervals, where $N=1$ ... $5$.  \label{fig:sigma-levels} }
\end{figure}

\begin{figure}
  \centering
  \resizebox{0.44\textwidth}{!}{\includegraphics{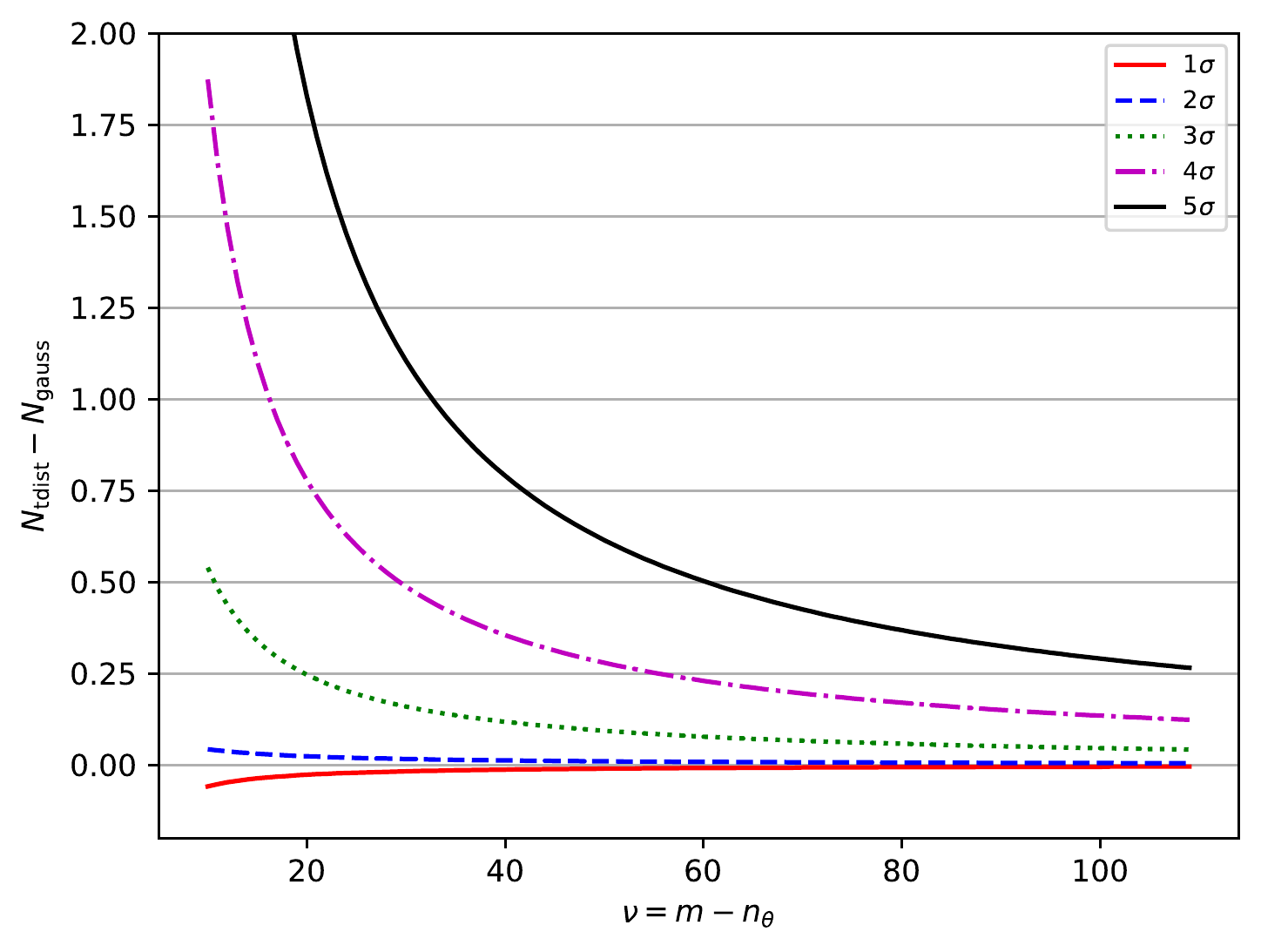}}
  \caption{The difference between the linear factors required to define credible intervals containing a fixed probability for t-distribution and Gaussian posteriors. Intervals are defined using the Gaussian probability within the $\pm N\sigma$ interval, where $N=1$ ... $5$. So, for example, tracing the $5\sigma$ curve (solid black line), we see that for $\nu=100$, \corr{to match the Gaussian $\pm 5\sigma$ coverage probability}, we would need to consider a $\pm5.29\sigma$ interval for the t-distribution. \label{fig:sigma-levels-N} }
\end{figure}

We now consider how the use of a multivariate t-distribution affects the interpretation of confidence intervals. Where credible intervals are derived directly from the posterior, for example, by considering the fraction of points within a given interval for a MCMC chain exploring a posterior volume, then the interpretation of results is correct whatever the form of the posterior. However, if one wants to define or interpret intervals based on $\pm N\sigma$ contours, then one needs to be careful when interpreting a posterior with t-distribution form, as explored in this Appendix.

As our favoured solution assumes a power law prior, the posterior, when written in terms of the model parameters for linear models, has a multivariate t-distribution form with degrees of freedom $\nu=m-n_\theta$. When marginalised over other parameters, the posterior probability for each model parameter has a form matching the student t-distribution for the parameter $\sqrt{(\nu-2)/\nu}{\bm\theta}_i/\sigma$. In general, the t-distribution has broader tails and a narrower core than the Gaussian distribution, matching the Gaussian distribution in the limit $\nu\to\infty$. The variance of the standard t-distribution is $\nu/(\nu-2)$, and so we need a broader range of integration to determine a $\pm N\sigma$ interval, integrating over $\pm N\sqrt{\nu/(\nu-2)}$ rather than $\pm N$ \corr{as with} a Gaussian for distribution with unit variance. The probabilities associated with credible intervals based on $\pm N\sigma$ are compared in Fig.~\ref{fig:sigma-levels}: the $\pm1\sigma$ credible interval is more probable for the t-distribution compared with the Gaussian distribution with the same variance. However, the tail probabilities are larger for the t-distribution than the Gaussian to fixed $\pm N\sigma$ limits for $N\ge2$. Fig.~\ref{fig:sigma-levels-N} instead shows the change in $N$ required to match tail probabilities from the t-distribution to those from the Gaussian distribution. For example, with a t-distribution posterior with $\nu=100$, one would need to define an interval based on the $\pm5.29\sigma$ threshold to match the inference (including tail probabilities) made from a $5\sigma$ result with a Gaussian posterior. We \corr{would therefore} need to integrate to larger intervals for the t-distribution to reduce the tail probabilities to match the Gaussian values for $N\ge2$. For smaller $\nu$ we need to integrate to larger intervals in $\sigma$.

\section{Analytic marginalisation for estimating the mean of data}  \label{app:marginalisation}

In this Appendix we outline the derivations that allow us to significantly speed up our Monte Carlo simulations fitting \corr{a single mean value $\bar{\mu}$ to $n_d$ correlated data values ${\bm x}_0$}, and ultimately would make them superfluous as we could perform all of the necessary calculations analytically. These are a special case of the derivation given in Section~\ref{sec:var_post_gauss} and are therefore not strictly necessary, but we include it as we feel that it gives insight into the problem being solved. To help with this, we first consider the more familiar case of a Gaussian posterior.

\subsection{Fitting the mean with a multivariate Gaussian posterior}  \label{sec:mean-gauss}

We start with the simple case of a Gaussian posterior. For this, we can use the standard definition of $\chi^2=-2\ln L$ for fitting
a mean $\bar{\mu}$ to data ${\bm x_0}$ with inverse covariance matrix $(hS)^{-1}$
\begin{equation}
  \chi^2 \equiv \sum_{ij} \left(({\bm
      x}_0)_i-\bar{\mu}\right)(hS)^{-1}_{ij}\left(({\bm x}_0)_j-\bar{\mu}\right)\,.
\end{equation}
Expanding, we can write
\begin{equation}
  \chi^2= h^{-1}\left[C_1-2C_2\bar{\mu}+C_3\bar{\mu}^2\right]\,,
\end{equation}
where
\begin{eqnarray}
  C_1&=&\sum_{ij}(x_0)_i S^{-1}_{ij}(x_0)_j\,,\\
  C_2&=&\sum_{ij} S^{-1}_{ij}(x_0)_j\,,\\
  C_3&=&\sum_{ij} S^{-1}_{ij}\,.
\end{eqnarray}
To align with the notation used elsewhere in this paper, we note that for this problem, the parameter $\theta=\bar{\mu}$, the model is $\mu_i=\bar{\mu}$, and we have $d\mu_i/d\theta=1$, $F_S=C_3$ and $F_S^{-1}=1/C_3$. \corr{The derivative} $\frac{d{\bm\mu}}{d{\bm\theta}}=U$, where the unit vector $U$ is a vector of 1's, and $C_2=U^TS^{-1}{\bm x}_0$. We now ``complete the square'' for the model dependent part of $\chi^2$
\begin{equation}
  -2C_2\bar{\mu}+C_3\bar{\mu}^2=C_3\left(\bar{\mu}-\frac{C_2}{C_3}\right)^2-\frac{C_2^2}{C_3}\,.
\end{equation}
We can then write the posterior as a Gaussian distribution around the MAP estimate
\begin{equation}
  f(\bar{\mu}|{\bm x}_0,S)\propto\exp\left[-\corr{\frac{C_3}{2h}}\left(\bar{\mu}-\frac{C_2}{C_3}\right)^2\right]\,.
\end{equation}
The mean, as derived from the posterior therefore has a Gaussian distribution, and the expected value for $\bar{\mu}$ and the variance can then be read off, $\langle \bar{\mu}\rangle_x=C_2/C_3$, and $\sigma^2=h/C_3$. As expected for a Gaussian posterior and a linear model, the MAP estimate matches the value given in Eq.~\ref{eq:thetahat}, and the expected model parameter variance integrating under the posterior matches the inverse of the Fisher matrix. \corr{So we see that} a Gaussian fit to the peak of the posterior also describes the results from the full distribution. 

\subsection{Fitting the mean with multivariate t-distribution posterior}  \label{sec:mean-tdist}

This section replicates Section~\ref{sec:var_post_gauss}, but now for the special case of fitting the mean to a set of data, as considered in Section~\ref{sec:MC_test}. We do this as we used these equations to speed-up the Monte Carlo runs presented in Section~\ref{sec:MC_test}, and in order to allow them to be used as an aide to understanding the derivation in Section~\ref{sec:var_post_gauss}. Consequently, we try to keep the layout and structure similar and make no apologies for replication. We only present the derivation for Gaussian distributed data.

Assuming that the posterior has a scale matrix $(n_s-1)/(m-n_d)S$, and degrees of freedom $\nu=m-n_d$, as in Eq.~\ref{eq:var_tdist_m} we can write the posterior where the model is a constant mean value $\bar{\mu}$
\begin{equation}  \label{eq:tdist-post-mean}
  f(\bar{\mu}|{\bm x}_0,S) \propto
  \left[ 1 + \frac{1}{n_s-1}\sum_{ij}\left(({\bm
        x}_0)_i-\bar{\mu}\right)S^{-1}_{ij}\left(({\bm x}_0)_j-\bar{\mu}\right)\right] ^{-\frac{m}{2}}\,.
\end{equation}
Expanding as in the Gaussian case, we have
\begin{equation}  \label{eq:tdist-in-Cs}
  f(\bar{\mu}|{\bm x}_0,S) \propto
  \left[ 1 + \frac{1}{n_s-1}\left(C_1-2C_2\bar{\mu}+C_3\bar{\mu}^2\right)\right] ^{-\frac{m}{2}}\,,
\end{equation}
and completing the square gives
\begin{equation}
  f(\bar{\mu}|{\bm x}_0,S) \propto
  \left[ 1 + \frac{1}{n_s-1}\left(C_1-\frac{C_2^2}{C_3}\right)+\frac{C_3}{n_s-1}\left(\bar{\mu}-\frac{C_2}{C_3}\right)^2\right] ^{-\frac{m}{2}}\,.
\end{equation}
We now define
\begin{equation}
  y = \sqrt{C_3}\left(\bar{\mu}-\frac{C_2}{C_3}\right)
  \left[1 + \frac{1}{n_s-1}\left(C_1-\frac{C_2^2}{C_3}\right)\right]^{-1/2}\left(\frac{n_s-1}{m-1}\right)^{-1/2}\,,
\end{equation}
so that
\begin{equation}
  f(\bar{\mu}|{\bm x}_0,S) \propto
  \left[ 1 + \frac{y^2}{m-1}\right]^{-\frac{m}{2}}\,.
\end{equation}
We see that $y$ is distributed with a t-distribution with $m-1$
degrees of freedom, such that the mean $\langle y\rangle=0$, and the
variance $\langle y^2\rangle=(m-1)/(m-3)$.

We can write $\bar{\mu}$ in the form $\bar{\mu}=ay+b$, which has the property that $\langle \bar{\mu}\rangle=\corr{a}\langle y\rangle+b$, and Var$(\bar{\mu})=a^2$Var$(y)$.
\begin{equation}
  \bar{\mu} = \frac{1}{\sqrt{C_3}}\left[1 +
    \frac{1}{n_s-1}\left(C_1-\frac{C_2^2}{C_3}\right)\right]^{1/2}
  \left(\frac{n_s-1}{m-1}\right)^{1/2}y+\frac{C_2}{C_3}\,.
\end{equation}
From this, we see that the distribution of $\bar{\mu}$ has mean
$\langle\bar{\mu}\rangle=C_2/C_3$, as expected given the discussion in Section~\ref{sec:DS}. The variance for any realisation is
\begin{equation} \label{eq:mean-var-full}
  \corr{\langle(\bar{\mu}-\langle\bar{\mu}\rangle)^2\rangle}
    = \frac{n_s-1}{m-3} \frac{1}{C_3}
    \left[1 + \frac{1}{n_s-1}\left(C_1-\frac{C_2^2}{C_3}\right)\right]\,,
\end{equation}
which matches Eq.~\ref{eq:var_post_givenxandS} for a fit to the mean. Thus, rather than numerically integrate under the posterior for any realisation of ${\bm x}_0$ and $S$, we can instead use this expression for the variance recovered. We have confirmed numerically that this result is correct, and that the variance depends on the data as given in this equation. Unlike for the Gaussian distribution, here the recovered variance depends on the value of ${\bm x}_0$ through $C_1$ and $C_2$. These terms do not cancel in general.

We can now consider the expected value, averaging over multiple sets of data, but using the same covariance matrix approximation $S$ to determine the posterior. In this case, $C_3$ is fixed, and we need to replace the terms $C_1$ and $C_2^2$ by the relevant expected values. Remembering that ${\bm x}$ was drawn from a Gaussian distribution with covariance $\Sigma$ \corr{and zero mean}, we have
\begin{eqnarray}  
  \langle C_1\rangle_x &=& \sum_{ij}[S^{-1}\Sigma]_{ij}\,,\label{eq:C1x}\\
  \langle C_2^2\rangle_x &=& \sum_{ij}[S^{-1}\Sigma S^{-1}]_{ij}\label{eq:C2x}\,.
\end{eqnarray}
This is the expected result for the variance recovered for many \corr{Gaussian distributed} realisations of the data ${\bm x}_0$. Eq.~\ref{eq:mean-var-full}, together with the expressions of Eqns.~\ref{eq:C1x} \&~\ref{eq:C2x}, allow us not to run Monte Carlo simulations for different data for the same covariance, as we can accurately predict the result using these equations.

To go one step further when finding analytic expressions for the Monte Carlo runs, we now need to find expressions for the relevant terms in Eq.~\ref{eq:mean-var-full}, now considering the expected values averaging over all possible covariance matrices $S$. We can do this using the expressions given in \corr{Section~\ref{sec:pert}} for the expansion of $S$ around the true matrix $\Sigma$. These give
\begin{eqnarray}
  \langle 1/C_3\rangle_S &=& [1+A+2B]h^{-1} F_\Sigma^{-1}\,,\\
  \langle C_1/C_3\rangle_S &=& [n_d+2B(n_d-1)] F_\Sigma^{-1}\,,\\
  \langle C_2^2/C_3^2\rangle_S &=& [1+B(n_d-1)] F_\Sigma^{-1}\,.
\end{eqnarray}
As expected, this final two equations match Eqns.~\ref{eq:pert-mid} \&~\ref{eq:pert-end} with $n_\theta=1$. For the first expression we write here the perturbative result rather than the exact form as used in Section~\ref{sec:var_post_tdist}. In terms of $n_\theta$, this is $\langle F_S^{-1}\rangle=h^{-1}[1+A+B(n_\theta+1)]F_\Sigma^{-1}$. Note that by using these expressions we would have removed any need to do the Monte Carlo simulations, as we have analytic expressions for all stages of the Monte Carlo runs being performed, albeit to second order in the covariance matrix approximation.

%

We can also consider how, for this case of fitting the mean to correlated data, we can derive an analytic expression for the scatter in recovered MAP estimates. To determine this, note from Eq.~\ref{eq:tdist-in-Cs} that the MAP estimate (obtained by taking the log and setting the derivative with respect to $\bar{\mu}$ to zero in the posterior) is $C_2/C_3$. From the definition of these quantities, $C_2/C_3=U^TS^{-1}{\bm x}_0/C_3$. Remembering that ${\bm x}_0$ are drawn from a Gaussian distribution with covariance $\Sigma$, we see that $C_2/C_3$ is also Gaussian distributed with zero mean and variance $U^TS^{-1}\Sigma S^{-1}U/C_3^2=\langle C_2^2/C_3^2\rangle_S$. Eq.~\ref{eq:var_dist} then shows that this matches the \citet{DS13} result.

A reader having reached this stage of the paper firstly needs congratulating, but also might well be asking why we need to run the Monte Carlo simulations presented in Section~\ref{sec:MC_test} at all given that we have analytically \corr{approximated} all of the results we will extract from those simulations. And they would be correct. However, we keep Fig.~\ref{fig:var} as it adds colour and \corr{confirms the validity of the approximations - using the Fisher matrix to determine confidence intervals from the posterior, and the second order expansions through which we estimated the impact of $S$.}


\bsp	
\label{lastpage}
\end{document}